\documentclass{sig-alternate}
\usepackage{mathptmx} % This is Times font

\usepackage{comment}

\usepackage{amsmath,amssymb,amsfonts}
\usepackage{algorithmic}
\usepackage{graphicx}
\usepackage{textcomp}
\usepackage{xcolor}

\usepackage{pdfpages}
\usepackage{listings}
\usepackage[hyphens]{url}
\usepackage{fancyhdr}
\usepackage[normalem]{ulem}
\usepackage[hyphens]{url}
\usepackage[sort,nocompress]{cite}
\usepackage[final]{microtype}
\usepackage[keeplastbox]{flushend}
\usepackage[bookmarks=true,breaklinks=true,letterpaper=true,colorlinks,linkcolor=black,citecolor=blue,urlcolor=black]{hyperref}

\usepackage{xcolor}
\usepackage{subcaption}
\usepackage{makecell}

\title{Timing Cache Accesses to Eliminate Side Channels in Shared Software}

\numberofauthors{2} %  in this sample file, there are a *total*
\author{
\alignauthor
Divya Ojha\\
       \affaddr{Department of Computer Science}\\
        \affaddr{University of Rochester}\\
       \email{dojha@cs.rochester.edu}
\alignauthor
Sandhya Dwarkadas \\
       \affaddr{Department of Computer Science}\\
        \affaddr{University of Rochester}\\
       \email{sandhya@cs.rochester.edu}
}

\begin{document}
\maketitle
\pagestyle{plain}

%%%%%% -- PAPER CONTENT STARTS-- %%%%%%%%

%    \input{sections/abstract.tex}
 \begin{abstract}

Timing side channels have been used to extract cryptographic keys and sensitive documents even from trusted enclaves. Specifically, cache side channels created by reuse of shared code or data in the memory hierarchy have been exploited by several known attacks, e.g., evict+reload for recovering an RSA key and Spectre variants for leaking speculatively loaded data.

In this paper, we present TimeCache, a cache design that incorporates knowledge of prior cache line access to eliminate cache side channels due to reuse of shared software (code and data). Our goal is to retain the benefits of a shared cache of allowing each process access to the entire cache and of cache occupancy by a single copy of shared memory. We achieve our goal by implementing per-process cache line visibility so that the processes do not benefit from cached data brought in by another process until they have incurred a corresponding miss penalty. Our design achieves low overhead by using a novel combination of timestamps and a hardware design to allow efficient parallel comparisons of the timestamps. The solution works at all the cache levels without the need to limit the number of security domains, and defends against an attacker process running on another core, same core, or another hyperthread.  

Our implementation in the gem5 simulator demonstrates that the system is able to defend against RSA key extraction. We evaluate performance using SPEC2006 and PARSEC and observe overhead due to first access delay to be 1.5\%. The overhead due to the security context bookkeeping is of the order of 0.3\%.

\end{abstract}

\section{Introduction}

Shared memory resources expose timing side channels that can reveal information even in the presence of security measures such as process isolation and enclave separation. Cache side channels leveraging shared memory have been shown capable of extracting cryptographic keys, sensitive documents, and data even from cryptographically secured enclaves~\cite{sgxpectre}. Several classes of cache side channel attacks and defenses have been developed in the literature~\cite{winter}.

In this paper, we focus on cache side channels created by the reuse of shared software (code and data) in the memory hierarchy. Shared software is an essential component to keeping system costs low. For instance, shared libraries (code) are an important optimization in modern computing systems to help keep the memory footprint low. Likewise, services providing access to large data stores result in data being shared across untrusted client requests. Access to the shared code or data leaves a footprint in the memory hierarchy, which has been exploited by several known attacks~\cite{PaaS}~\cite{games}~\cite{llcpractical}~\cite{flushreload}~\cite{flushflush}~\cite{sgxpectre}~\cite{spectre}.

A typical cache side channel attack when sharing software involves evicting the shared data (e.g., code from a shared library) from the cache hierarchy and re-accessing it after the victim's execution (using evict+reload or flush+reload~\cite{flushreload}). A fast re-access is indicative of an access to the shared location by the victim. If the shared library access is indexed by a secret data, the attacker can infer the victim's secret. This attack model is used in attacks to leak cryptographic keys~\cite{flushreload}, in Spectre I, Spectre II~\cite{spectre}, NetSpectre~\cite{netspectre},  in cross-tenant attacks to leak data in clouds providing Platform-as-a-service~\cite{PaaS}, and in discovering key strokes~\cite{unveiling}. 

There is another class of cache side channel attacks that do not require shared memory and is not the focus of this work. These attacks are referred to as contention or conflict-based side channel attacks. Contention attacks are mitigated using randomizing caches as in  CEASER~\cite{ceaser}, CEASER-S~\cite{ceasers}, and  ScatterCache~\cite{crosscoreAttack},  or using very efficient multiple hashing techniques like RPCache~\cite{newcache}. However, these techniques are unable to prevent reuse attacks in shared memory. TimeCache can work in conjunction with these techniques to provide a holistic defense.

Reuse attacks on shared memory are more precise and a handy tool for constructing more sophisticated attacks. They are less noisy and are a preferred covert channel for leaking speculatively loaded data~\cite{ridl,fallout,spectre}. Preventing reuse attacks on shared memory will not only make the attacker's work difficult but also allow system providers to deploy deduplication or copy-on-write sharing (e.g., unix-style process fork operations or Docker-style containers) for increased performance and reduced space utilization. Deduplication evaluation in the literature shows that its use can %save 80\% or more in terms of space~\cite{dedup1}
%. Singleton uses KSM (kernel same-page merging) for deduplication,
reduce memory needs by a factor of 2-4x~\cite{dedup1,singleton} and increase performance by up to 40\%~\cite{singleton}. 

%SD: are these simulations or on real machines?
Existing solutions for reuse attacks partition the cache~\cite{secdcp} ~\cite{static} ~\cite{dawg} or implement constant time algorithms, both resulting in increased latency. Partitioning has been seen to be associated with higher overheads due to both reduction in the effective cache size for individual processes, and due to potential aliasing of the shared memory in the cache, depending on the system design. Partitioning techniques also might have restrictions on the number of supported security domains~\cite{dawg,newcache}. 
%SD: We need to discuss and clean this up.
For instance, DAWG~\cite{dawg} supports only 16 security domains at a time.  
%or assume processes as safe and unsafe~\cite{secdcp}, which limits the scope of the solution. 
%SD: did not see the need for or relevance of the above elaboration
Some other solutions protect accesses only to the LLC~\cite{catalyst}~\cite{sharp}.
%SD by doing what?

In this work, we design and evaluate a low-overhead hardware-software solution to defend against reuse attacks. Our goal is to retain the benefits of a shared cache of allowing each process access to the entire cache and of cache occupancy by a single copy of shared software.  We protect every level of cache without limiting the number of supported security domains. Our key insight is recognizing the importance of the attacker's \emph{first access} to the data after a victim has brought the data into cache. We want to ensure that the first access by a process to any cache line loaded by a different process is a miss. 

TimeCache creates a ``per-process view'' of cache occupancy by delaying (treating as a miss) the \emph{first access} to a cache line (i.e., a cache hit).
%that has been brought into the cache by a different process. 
%Accesses beyond the first will be serviced as a hit. 
The delayed first access gives the process the illusion of timing isolation from other processes. While there is shared data between processes, the isolation is achieved by giving every process the impression that data is brought into the cache by its own access. This approach breaks the fundamental premise of a reuse attack  using shared software. The reduction in the performance due to such delay can be considered elemental to the design of a secure cache. It avoids a potential $O(n)$ space consumption for $n$ processes sharing data in a partitioned cache. The delay is incurred only when data is evicted and reloaded, so that the steady-state in-cache sharing is unaffected. 
As a consequence of this defense, systems can choose to deploy memory deduplication techniques to reduce memory footprint~\cite{dedup1,dedup2} without the fear of creating an avenue for cache side channels through a shared software stack.
%SD need to delve into the citations above

We propose and evaluate a timestamp-based cache access management system that combines novel hardware and software support to provide timing isolation. 
%In hardware, each cache line is augmented with a ``load-time'' timestamp and a bit per hardware context to mark the caching behaviour.
%This bit is checked upon a cache hit and the request is serviced if the bit is set; otherwise, a miss penalty is incurred and a request is sent down the memory hierarchy. 
In software the process's caching context is saved (and restored) along with the ``context-switch'' timestamp at a context switch. A novel hardware-implemented  \emph{bit-serial} comparison logic allows fast parallel timestamp comparisons, which is used to update the stale caching context.

The security and performance of TimeCache on the gem5 simulator~\cite{gem} demonstrates its effectiveness against attacks using microbenchmarks and an RSA attack. Our defense is able to prevent the classic RSA attack used to demonstrate {\tt flush+reload} attacks. Performance evaluation using SPEC2006 and PARSEC shows an average overhead of 1.5\% and 1.2\%, due to the delayed accesses. The overhead due to the security context bookkeeping is about 0.24\%.

This paper makes the following key contributions:
\begin{itemize}
    \item Preventing reuse attack while allowing entire cache access for each process and maintaining a single cached copy.
    \item The insight that disallowing the \emph{first access} by a process to a cache line from experiencing a cache hit when the cache line has been brought into the cache by a different process, is sufficient to prevent cache side channel attacks via reuse of shared software.
    \item A timestamp-based solution to updating a per-process view of cache line occupancy across context switches to prevent such attacks.
    \item A fast \emph{bit-serial} comparison logic to compare timestamps for all the cache lines simultaneously.
    \item A simulation-based evaluation of the potential overheads of timing isolation and a demonstration that our proposed solution prevents real-world attacks.
 \end{itemize}

%The following sections describe the design, proposed implementation, performance, and security analysis of our solution.
\section{Background}
\subsection{Cache Side Channels}

Information leaked as a result of shared cache utilization is collectively referred to as \emph{cache side channels}. Mechanisms to exploit cache side channels were first exposed as early as in 1992~\cite{wei-ming92}, and
%SD: is citation 13 a single page?
different classes of attacks relying on cache access timing have been developed since then. 

\begin{figure}[h]
\caption{Reuse and Contention Attacks}
\label{cratt}
\centering
\includegraphics[width=0.48\textwidth]{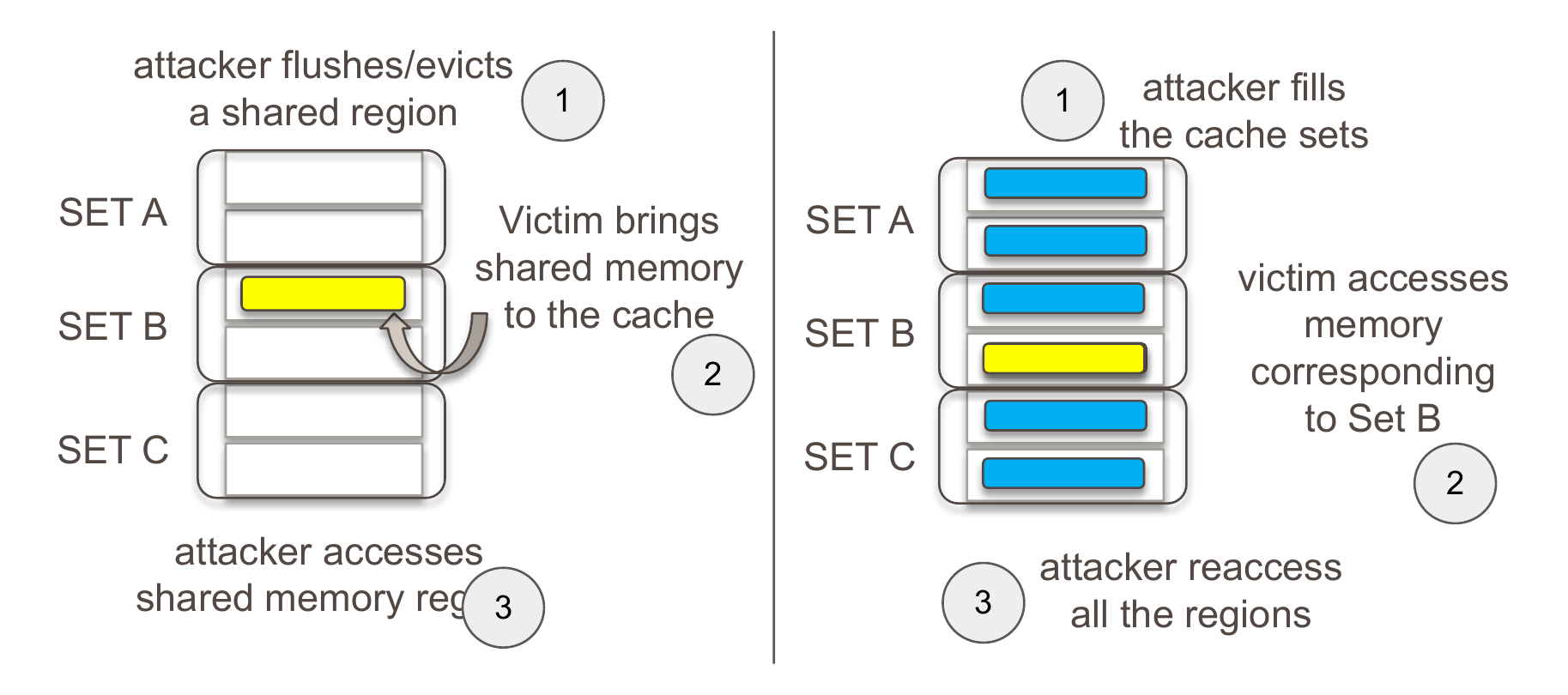}
\end{figure}

Two types of information leak are possible depending on whether or not there is shared software between the attacker and the victim. With no shared software between the attacker and the victim, the attacker can only learn the cache set accessed by the victim using a ``Prime+Probe''~\cite{primeprobe} style of attack, commonly referred to as contention-based attack. In the presence of shared software, an attacker can learn the line accessed by the victim using an ``evict+reload'' or ``flush+reload'' style attack~\cite{flushreload,templateattack}. Figure~\ref{cratt} depicts both the attacks. 
Defenses against ``prime+probe'' attacks such as caches using randomized placement~\cite{ceaser}~\cite{newcache-micro-2016} are not effective against ``evict+reload'' or ``flush+reload'' style attacks. 
The latter is a low-noise, high-bandwidth, and more efficient form of attack. This work addresses this second style of attacks.
%, since they depend on shared libraries to leak application secret. They also form a gadget usable for constructing attacks like Spectre.

\subsection{Shared Software Attacks}

Shared libraries have commonly used subroutines, which can be mapped directly into a process's user-level address space. The same physical memory may be mapped into different processes' virtual address spaces. Shared libraries help reduce memory footprint and improve memory hierarchy efficiency. However, they create the potential for leaked memory access patterns through cache side channels, whether due to access to the shared code or to shared data.

Side channels exist due to shared hardware and software, and improve in precision in the presence of shared software. Side channels using shared software were earlier assumed to affect only cryptographic routines. The advent of more recent attacks have shown that they are a handy gadget for more sophisticated attacks like Spectre~\cite{spectre,netspectre,spectreRSB}. They are also capable of leaking keystrokes from another process~\cite{unveiling}, leaking passwords in cloud environments such as an Amazon EC2 server, and leaking data across Virtual Machines~\cite{PaaS}. 

\subsection{Prior Solutions}
There have been a number of defenses around cache side channels but the strict performance requirement of caching structures has remained a challenge. One class of defenses mitigate only contention-based attacks (e.g., randomizing caches~\cite{ceaser, ceasers, scattercache}, SHARP~\cite{sharp}, RPCache~\cite{newcache}). Defenses that include reuse attacks either resort to cache  partitioning~\cite{newcache,dong2018shielding} or implement constant time algorithms~\cite{rane15raccoon}. Partitioning defends against both reuse and contention attacks but reduces the effective cache available for each process execution, effectively reducing performance. Constant time algorithms, likewise, incur a significant performance penalty~\cite{rane15raccoon,rane16,ghostrider}. Some defenses restrict the number of possible security domains~\cite{newcache,dawg}. Others work only for the last level cache~\cite{winter}.  

In terms of detecting the presence of side channels, hardware description languages that check for information flow in the hardware specification~\cite{hdl} help to detect the existence of side channels and help specify security labels in the processor for building side channel resistant processors. Likewise, Checkmate uses relational model finding to detect the presence of side channels in processor specification~\cite{checkmate}.

\section{Threat Model}

\begin{figure}[b]
\caption{Reuse Attack in Cache}
\label{attack}
\centering
\includegraphics[width=0.48\textwidth]{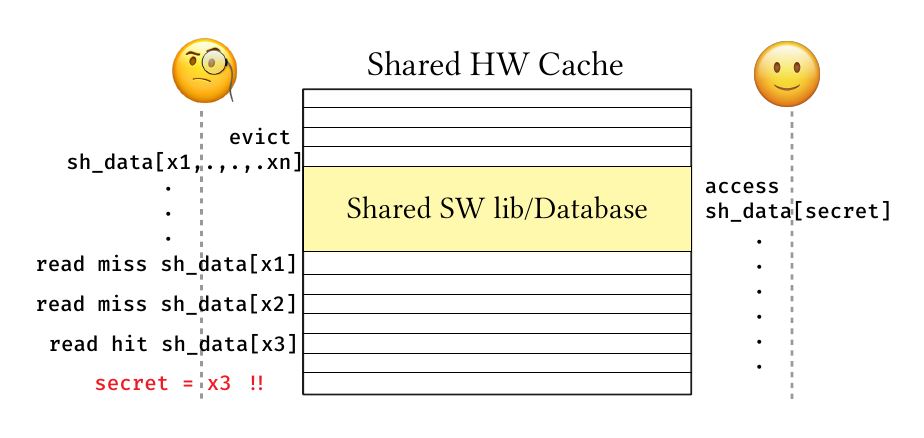}
\end{figure}

The threat model under consideration has a separate attacker and  victim process sharing some software stack in addition to sharing caches. They could be running simultaneously on the same (hyperthreaded) or different cores, or interleaved in time, and the attack can be conducted from any level of the cache. Figure~\ref{attack} shows a reuse attack when accessing the shared software stack in a cache shared between attacker and victim. The attack has the following sequence:
\begin{enumerate}

    \item The attacker and victim share software and a hardware cache. The access to the shared software is dependent on or is indexed using the victim's secret data.
    \item The attacker evicts a shared location from the cache hierarchy.
    \item It then waits for the victim's execution.
    \item The attacker subsequently reloads the same shared location and determines that the shared location was also accessed by the victim if it hits in the cache, determined by timing the access.

\end{enumerate}

This attack model is self-sufficient in the sense that it has been demonstrated to be capable of leaking RSA keys when using the GnuPG shared library~\cite{flushreload,games}. 
%SD: missing citation above
It is also a low-noise, high-bandwidth tool for building more sophisticated attacks like Spectre-I \& II~\cite{spectre}, SpectreRSB~\cite{spectreRSB}, and Netspectre~\cite{netspectre}. It is a preferred covert channel for these more recent attacks for leaking shared library access patterns. Other contention-based covert channels could also be used for similar purpose but are not as precise and hence would make the attack difficult.

\section{TimeCache Design}
\label{sec:design}

TimeCache eliminates reuse cache timing side channels by implementing techniques to allow per-process cache line visibility of a shared cache line. Unlike partitioning-based approaches, TimeCache does not use isolation in space; rather, TimeCache ensures that accesses to a shared cache line by different processes are isolated in timing. 
%\textcolor{pink}{Every process should see its \emph{first access} to a resident cache line that it did not bring in with a delay equivalent to a miss in the cache. Once a cache line has been accessed by a process, its subsequent accesses to the data from the same cache line are allowed to go through as a hit even after context switches (presuming the data has not been evicted)}. 
This allows different processes to share access to the same cache line without revealing to one another if the cache line was made available in the cache by another process. 
Compared to solutions that rely on cache partitioning (example, Intel's cache allocation technology~\cite{catalyst} in~\cite{dong2018shielding}),  this approach does not restrict the size of usable cache for any process. It is also applicable to any level of the cache hierarchy from L1 to LLC. 

%\subsection{Goal \& Insights}
%The potential scope of cache side channel exploits include those devised in Spectre-style attacks based on speculation~\cite{spectre} and is not limited to specific application domains such as cryptography. 
%Hence,  domain- or attack-specific solutions are not sufficient.

%The goal of this work is to design a cache that allows processes to reap the benefits of sharing both cache and data without leaking access pattern information via timing side channels. 
%Preventing information leak requires identifying and slowing down the first access to data present in the cache but brought in by a different process. 
% As a consequence of this design, systems can choose to deploy memory deduplication techniques to reduce memory footprint~\cite{dedup1},~\cite{dedup2} without fear of creating an avenue for a reuse-based covert channel attack.

%In the following subsections, we introduce the notion of the \emph{first access} and then describe the hardware needed to perform the required checks on accesses for allowing fast or slow response with low overhead. We also describe the software changes required with the new hardware to maintain the caching benefits across preemption.

\subsection{First Access}
\label{sec:firstacc}

A process's \emph{first access} refers to the first time it accesses a resident cache line that was brought into the cache by another process. A resident cache line can experience as many \emph{first access} misses as the number of processes accessing it (minus one for the initial cache line fill). If a cache line is evicted and later brought back into the cache by a process, all other processes accessing the cache line at a later time will experience a \emph{first access} miss.

The importance of the \emph{first access} lies in the construct of the attack. If the attacker times its \emph{first access} after evicting a data from the cache hierarchy and is able to detect a cache hit, the attacker is able to infer the victim's memory access patterns. Beyond the \emph{first access}, a fast access or a cache hit does not provide any clue about the data access pattern of another process. Using this key observation, TimeCache enables space sharing (avoid partitioning) by enforcing misses on \emph{first access} to provide timing isolation. 

In the baseline cache design, processes experience delays in access due to cold, capacity, conflict, and coherence misses, whether due to its own actions or due to those of other processes. Critically, it may also experience hits in the cache due to data brought in because of another process's access. It is this latter timing side channel that TimeCache targets. In TimeCache, we create a new kind of miss: a \emph{first access} miss.

\subsection{Distinguishing First Accesses}

\begin{figure}[b]
\caption{TimeCache Hardware}
\label{diag:cachedef}
\centering
\includegraphics[width=0.47\textwidth]{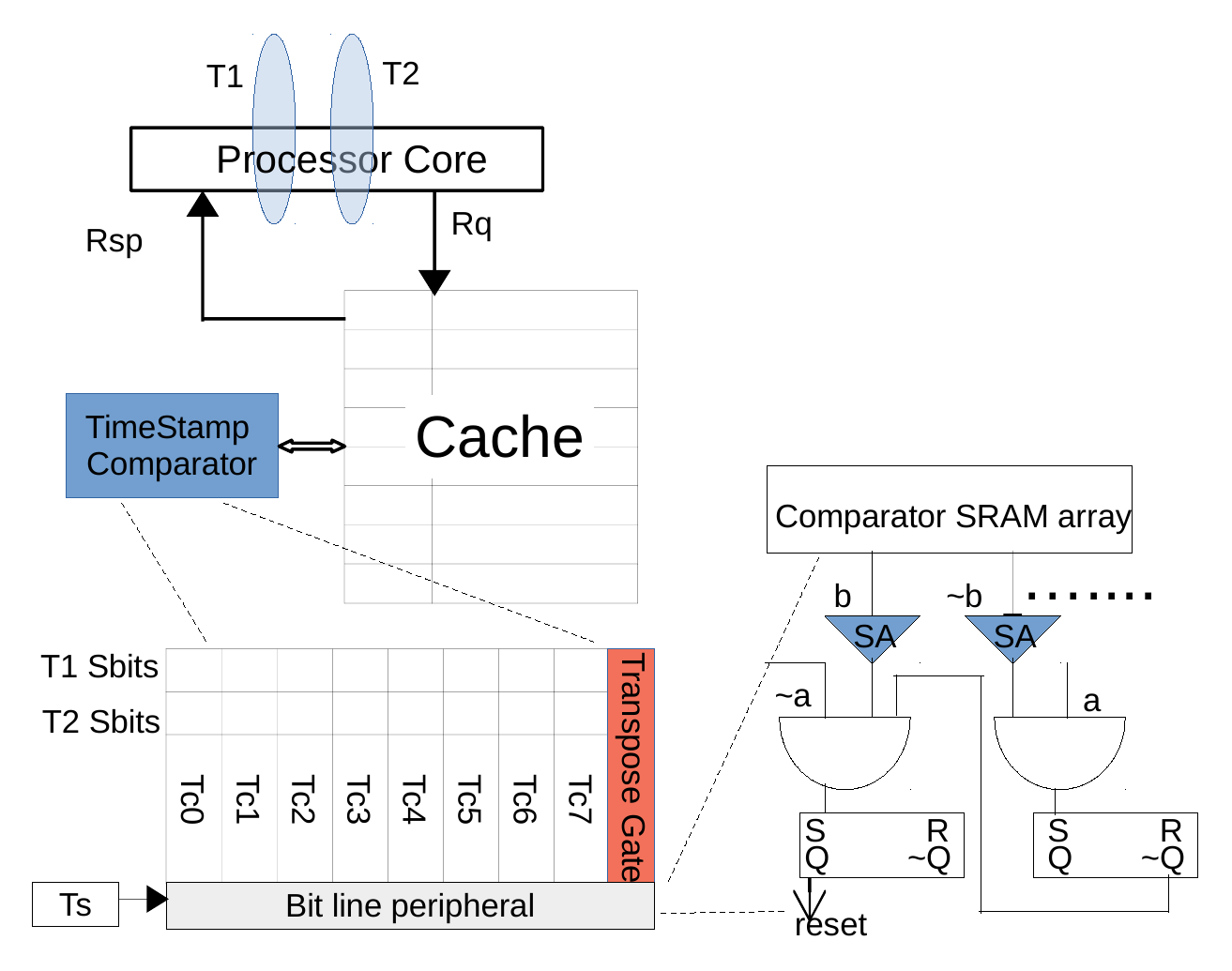}
\end{figure}

TimeCache is based on the observation that the attack under consideration exploits the caching benefits due to another process. 
%The attacker evicts the shared memory and expects its subsequent access to be fast if the victim has accessed the same shared memory location. 
Hence, we propose to identify and delay the \emph{first access} of a process, so the attacker is unable to infer whether the memory was cached beforehand. 
%Subsequent accesses by the process proceed as a cache hit. 
Figure~\ref{diag:cachedef} provides an overview of the hardware modifications. 

%\textcolor{pink}{In order to maintain per-process cache line visibility information and identify first accesses, an additional bit (security bit or \emph{s-bit}) per hardware context is added to each cache line. The \emph{s-bit} represents whether the cache line has already been accessed by the context.} 
Whether a cache line has already been accessed by the context is represented by a security bit or \emph{s-bit}. Figure~\ref{fig:s-bit-restore} outlines the modified sequence of operations for a cache access under TimeCache. When a cache line is brought into the cache (FillCacheLine) on a miss, the \emph{s-bit} for the loading context is set and the \emph{s-bits} for all other hardware contexts are reset. For a cache hit, the \emph{s-bit} of the accessing context is checked. When the \emph{s-bit} is set, the access is allowed to proceed as a hit. Otherwise the access is recognized as a \emph{first access}, treated as a miss, and the context's \emph{s-bit} is set so that the future accesses can proceed as a hit.  \emph{First access} misses are handled by sending the request down the memory hierarchy but not filling ({\tt no\_FillCacheLine}) the cache with the received data, as the data in the cache is the most recent copy. This mechanism is implemented at every level of cache in the memory hierarchy. 

\subsection{Handling Context Switches}

\emph{S-bits} represent the caching footprint of a hardware context and are specific to the process executing in the context. At the time of a context switch, \emph{s-bits} must be carefully managed in order to ensure that they are neither stale nor contain information on the caching context of a different process. 

In order to retain caching behavior across context switches and still provide timing isolation, software saves the \emph{s-bits} for a process along with current time as its  ``context-switch'' timestamp (\emph{Ts}) at a context switch. 
Additionally, each cache line is augmented with a ``load-time'' timestamp (\emph{Tc}) to store the time at which the line was loaded.

TimeCache ensures timing isolation at the time of a context switch using a combination of software and hardware. Software restores the \emph{s-bits} saved from the last time the process executed to the cache in the corresponding hardware context when a process resumes execution. Hardware ensures that the stale \emph{s-bits} (reflecting the state of the cache when the process last executed) are updated to reflect the current cache content.

The \emph{s-bit} save and restore can be done by any trusted computing base library at the time of context switch. 
%SD: not sure what you mean by library here. 
%Trust in the operating system (OS) is not a requirement and any secure software can keep the integrity and confidentiality of the \emph{sbits}. 
In our implementation, we  allow the OS to save and restore the process-specific \emph{s-bits}.
%In case of SGX enclave, the enclave setup routine could be do the \emph{sbit} book keeping. 

\begin{figure}
\caption{TimeCache Access}
\label{fig:s-bit-restore}
{\tt
\begin{lstlisting}
    /*memory access*/
    if (cache_miss)
        request_mem;
        FillCacheLine;
        s-bit=1 (for this context);
    else if (cache_hit)&(s-bit==0)
        request_mem;
        no_FillCacheLine;
        s-bit=1 (for this context);
    /*cache line replacement*/
    if (line_evicted)
        for_all_threads:
            s-bits = 0;
\end{lstlisting}
}

\end{figure}

%When the software invokes an \emph{s-bit} restore at the time of a context switch, \emph{s-bits} for cache lines whose \emph{Tc} is younger than \emph{Ts} must be reset, as this indicates that the content of the cache line was reloaded since the last time the processes was executing. A novel \emph{bit-serial}, timestamp-parallel comparison logic allows fast parallel \emph{Tc} and \emph{Ts} timestamp comparisons. The details of the implementation are described in Section~\ref{sec:impl}. 

Figure~\ref{diag:cachedef}, shows a cache consisting of 8 cache lines accessed by two hardware contexts. The hardware support added to a conventional cache is as follows:

\begin{itemize}
    \item A bit-serial, timestamp-parallel comparison logic with transpose gate and bitline peripherals, to compare timestamps efficiently.
    \item A per cache line, per hardware context bit (security bit or \emph{s-bit}).
    \item A shift register to hold Ts, the timestamp indicating the time when the process last executed, for the process about to begin execution due to a context switch.
\end{itemize}

The ability to save, restore and update the caching contextallows TimeCache to enjoy fast access as long as the data is not evicted from the cache.
%TimeCache leverages the benefits of caching across context switches as long as a cache line is not evicted. Two processes running in an interleaved fashion and accessing the same memory location will continue to enjoy fast access as long as the data is not evicted from the cache. After each eviction, each process will individually see a delayed \emph{first access}. Hence, 
Our design leverages locality across context switches while providing timing isolation, something that cannot be achieved by simply flushing the cache on a context switch. The mechanism described here is a processor feature which can be turned off if the processes are trusted and cache attacks are not a concern.

\section{Implementation}
\label{sec:impl}
The following subsections describe the implementation details of each hardware modification and the software support required for the defense.

\subsection{First Access Delay Mechanism}
On a traditional cache access, the requested data is returned to the processor if a tag and state lookup succeeds. Otherwise, the access incurs a miss and the request is passed on to the next level in the memory hierarchy. With our cache design, the \emph{s-bit} for the cache line is checked in addition to the state and tag bits. An access is considered a hit only if in addition to the above, the \emph{s-bit} of the cache line is set, in which case data is returned to the processor from the cache.  

\begin{figure}[]
\caption{Cache Access Flow Chart}
\label{flw}
\centering
\includegraphics[width=0.45\textwidth]{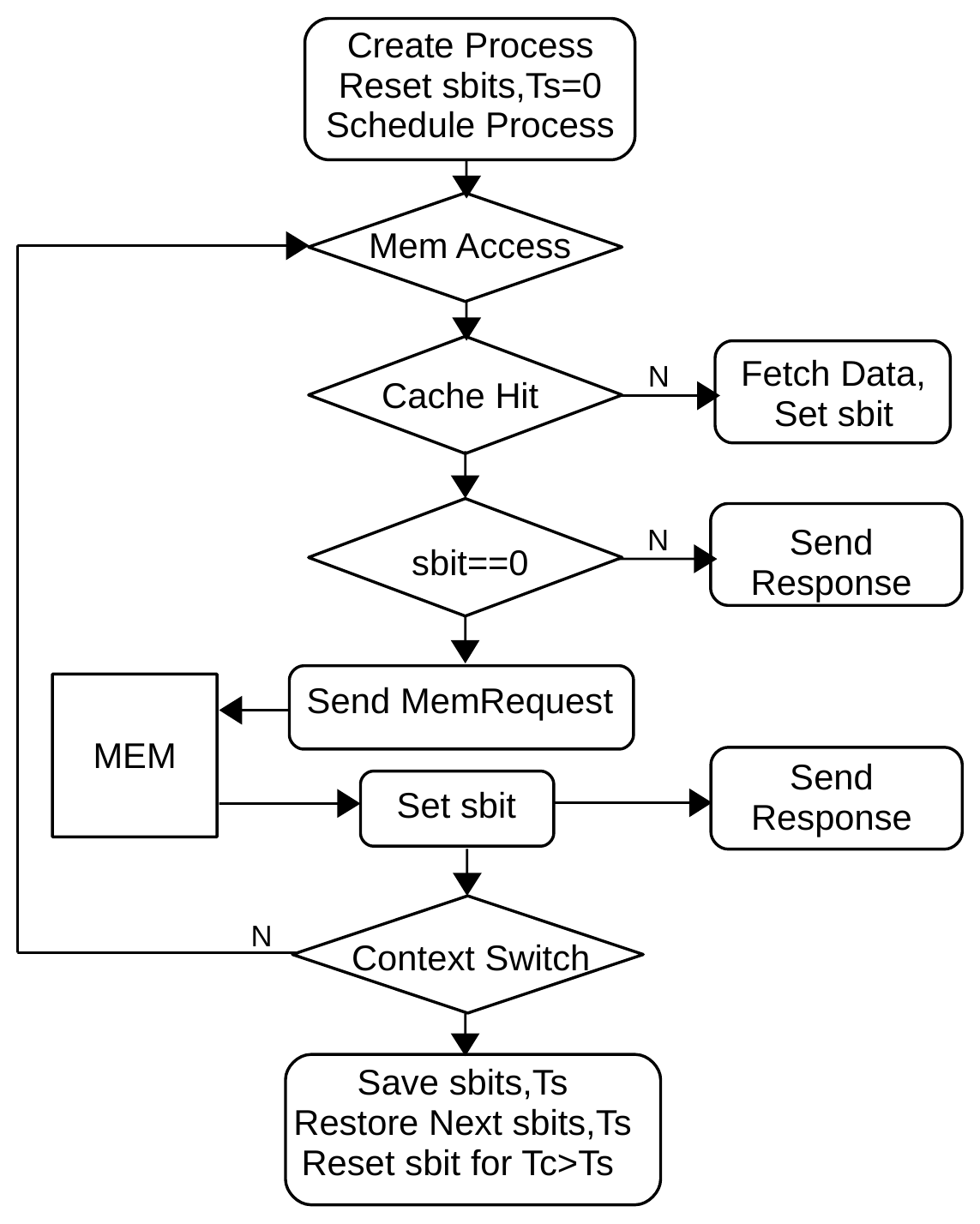}
\end{figure}

The flow chart in Figure~\ref{flw} represents the actions taken by TimeCache on behalf of a process during creation and execution.
Software saves and restores the \emph{s-bits} for a process executing on a hardware context at the time of a context switch. Additionally, software maintains  \emph{Ts} for each process, which is the time the process was most recently preempted. A newly created process has both \emph{Ts} and \emph{s-bits} reset when it is scheduled for the first time.

\emph{s-bit} state is changed as follows:
\begin{itemize}
    \item Reset when a process is first initialized
    \item Reset by the timestamp comparison logic when the process is scheduled
    \item Reset when a cache line is evicted or invalidated
    \item Set for the requesting hardware context when a cache line is loaded; reset on all other hardware contexts
    \item Set for the requesting hardware context after a first access to a previously loaded cache line
\end{itemize}

A reset \emph{s-bit} indicates that the current process has not accessed the cache line. If the \emph{s-bit} is reset, the response to the processor is delayed by sending the request down the memory hierarchy. Once the response is received, the received data is discarded, and the data in the cache line is forwarded to the processor. The \emph{s-bit} is set to ensure that future accesses to the cache line by the process do not result in additional traffic and are treated without additional delay. 

The rationale behind sending a request down the memory hierarchy even when the data is available in the cache is to make the \emph{first access} see a response latency equivalent to the variable access latency it would have incurred on a miss. It is possible that a context's \emph{s-bit} is reset in a higher-level (closer to the processor) cache but set in a lower-level cache due to its larger capacity. Sending a request down the memory hierarchy ensures that if the requested data is available in a lower-level cache and has the \emph{s-bit} set, the request is serviced with the lower cache response latency. The data received in the response is, however, discarded, as the cache has the most recent copy of the data. 

When a cache line is evicted or invalidated, all \emph{s-bits} are reset. 
When a cache line is loaded, the \emph{s-bit} for the hardware context loading the line is set;  the \emph{s-bits} for all other hardware contexts sharing the cache remain reset. 

On a context switch, hardware compares \emph{Tc} for each cache line against \emph{Ts} (loaded into a special register by software) for the context being resumed; the s-bits for lines that have \emph{Tc} greater than \emph{Ts} are reset to enforce delayed first access.

The \emph{s-bits} save and restore is done only at the context switch time, all the accesses thereafter proceed with an additional 1 bit lookup which is carried out in parallel with access to the cache data array. If the \emph{s-bit} is not set, the access results in a miss. This is unlike the conventional caches, where a response is sent back to the processor if the data is cached.

\subsection{Per-Process s-bits Copy and Update}
%The \emph{sbits} are indicative of whether or not a cache line has been accessed by a process before. 
The \emph{s-bits} are saved and restored on a context switch to preserve caching benefits across context switches. If the \emph{s-bits} were not saved and instead reset on every context switch, this would be equivalent to flushing the cache on every context switch, which can impact performance heavily~\cite{dong2018shielding}.

The number of 64-byte (cache line size) memory accesses required to save or restore \emph{s-bits} is dependent on the cache size. A small 64KB L1 cache requires only 2 64-byte memory accesses, while a larger 8MB L3 cache requires 256 64-byte memory accesses. 

Restored \emph{s-bits} cannot be used as is since they are stale and need to be updated based on any changes in the cache. If a cache line is evicted while a process is preempted, its corresponding saved \emph{s-bit} in memory will not be up-to-date. To update the \emph{s-bits} for cache lines that might have been evicted, invalidated, or reloaded when the process was preempted, we use the \emph{Ts} timestamp. \emph{Ts} indicates the last time the \emph{s-bits} were brought up-to-date, so any cache lines loaded after that time would not have been accessed by the process. When a process resumes execution, its restored \emph{Ts} is compared with the \emph{Tc} of every cache line in parallel, and the \emph{s-bits} for all cache lines with \emph{Tc} greater than \emph{Ts} are reset. Timestamp comparisons are triggered \emph{only} at the time of a context switch and prior to resuming a process. Subsequent accesses need no comparison since the \emph{s-bits} now contain the necessary information. 
Comparing timestamps serially for all cache lines at the time of a context switch can consume a significant number of cycles. 
We discuss the comparison of \emph{Tc} and \emph{Ts}, and the mechanism for updating large arrays of \emph{s-bits} in constant time (proportional to the number of \emph{Tc} bits) in the following subsection.

\subsection{Bit-Serial, Timestamp-Parallel Comparison of Timestamps}

A regular data access from memory is \emph{bit-parallel}, i.e, all the bits of a word are accessed at a time. Accessing SRAM arrays in \emph{bit-parallel} fashion implies that the time required would be proportional to the number of cache lines. In order to perform parallel comparisons of cache line timestamps (Tc) and Ts, we store the per cache line \emph{Tc} timestamps along with the cache line's \emph{s-bits} in an SRAM array in a transposed fashion, similar to that proposed in the neural cache work~\cite{neural}. The result is computation performed in a \emph{bit-serial}~\cite{bitserial} and word-parallel (timestamp-parallel) manner. 
%Rather than create separate SRAM arrays for bit-line computation (which we do not need) and use a separate transpose memory unit, we store the timestamps as the transpose array. 

\subsubsection{Transpose Interface}
\begin{figure}[tb]
\caption{Transpose SRAM Array for Timestamps}
\label{tx}
\centering
\includegraphics[width=0.22\textwidth]{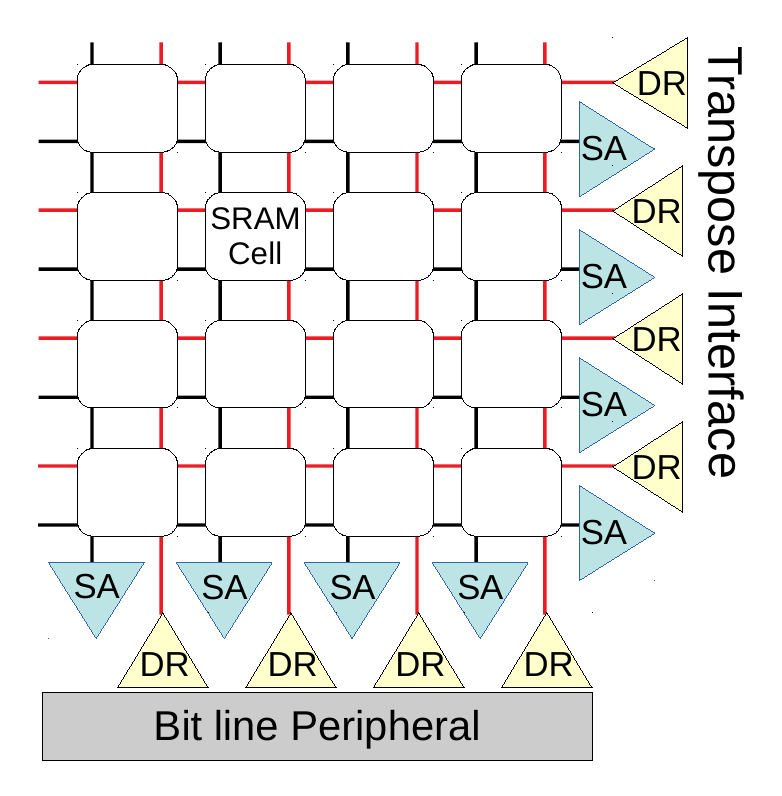}
\end{figure}

The transpose memory unit~\cite{neural} uses 8-T bit cells and two sets of sense amps and drivers to access data in both regular and transposed modes. While access times will be higher compared to a 6-T SRAM cell, accesses can be made in parallel with the much larger cache data arrays. 
Figure~\ref{tx} shows the timestamp array and comparison logic, constructed with the 8-T multi-access SRAM cells. The `transpose interface' is used for the regular operation of the cache, which is when timestamps are updated and \emph{s-bits} of other contexts are reset, or an \emph{s-bit} needs to be looked up or set. The `regular' bit-line peripheral interface is used for \emph{s-bit} saves and restores, as well as for parallel timestamp comparisons and \emph{s-bit} resets. 
%SD - added the need form sbit saves and restores

After the process-specific \emph{s-bits} are loaded into the SRAM array in the \emph{s-bits} for the corresponding hardware context, they need to be updated with the information about the cache lines that have been evicted while the process was preempted. This is done by comparing the \emph{Tc} and the restored Ts.  The transposed timestamps allow a \emph{bit-serial} and timestamp-parallel comparison, taking time linear in the number of bits in the timestamp (32 in our experiments). The timestamp rollovers are discussed in Section~\ref{rollover}.
 The  logic required for the timestamp comparisons and reset of \emph{s-bits} is shown in Figure~\ref{peripheral}.

\subsubsection{Bit-Serial Comparison Logic}
\label{comparator}
Bit-serial computation allows us to simplify the comparison logic. 
The greater of two unsigned integers can be determined by comparing their bits sequentially starting from the MSB (most significant bit): one of the two numbers can be declared as larger when the first bit that differs is encountered: the larger number will have the bit set in its binary representation where the other number has the bit set to 0. We codify the above algorithm in the following scheme iterating from the MSB:
\begin{itemize}
\item If the bit position under consideration has a 1 for only one of the two numbers, that number can be marked as greater and the comparison is complete. This behavior can be checked by performing an {\tt xor} of the two bits.
\item If the bit position under consideration has a 0 for both the numbers, the next bit position is considered.
\item If the bit position under consideration has a 1 for both the numbers, the next bit position is considered.
\end{itemize}

For instance, the greater of the two numbers `1100' and `0101' can be determined as the first number `1100' by looking at the MSB.

\begin{figure}[tb]
\caption{Bit-Line Peripheral}
\label{peripheral}
\centering
\includegraphics[width=0.2\textwidth]{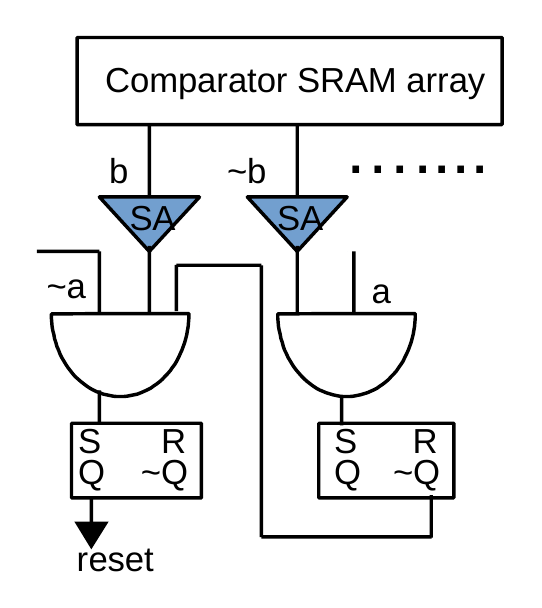}
\end{figure}

Ts is loaded into a shift register. For each of 32 iterations (the size of our \emph{Tc} timestamp), the \emph{Tc} timestamps are read from the SRAM array 1 bit at a time using the `regular' bit-serial peripheral interface, at the same time as the shift register is shifted left to feed the comparison logic. 

\begin{itemize}
    \item If \emph{Tc[i]} is 0 and Ts[i] is 1, $Tc < Ts$, the \emph{s-bits} need not be updated and the comparison should stop. We latch this output and use it to ignore further bit comparisons.
    \item If \emph{Tc[i]} is 1 and Ts[i] is 0, $Tc > Ts$, i.e., the cache line is newer than the Ts. The bit-line peripheral latches a `1' and the latch output is used as the reset for the \emph{s-bit}.
\end{itemize}
Figure~\ref{peripheral} shows the peripheral circuit attached to each SRAM bitline. It requires 2 SR latches, which are reset prior to initiating the timestamp comparisons, and 2 3-input {\tt and} gates for the comparison operation, with the \emph{Tc} bit being fed to `b' and the \emph{Ts} bit to  `a'.

The comparison should stop if the \emph{Tc} is determined to be smaller than Ts, which is the result of the {\tt and} gate on the right. To ignore further bit comparison, the result is latched using an S-R latch, and ~Q is fed to the {\tt and} gate on the left.  

At the end of the 32 iterations, if it is determined that $Tc > Ts$, as latched in the left-hand S-R latch, the bitline drivers for which the S-R latch has been set, and the wordline for the \emph{s-bit} corresponding to the hardware context, are enabled, to write a 0 into the \emph{s-bits}.

\section{Evaluation}
We implemented TimeCache in the gem5 cycle-accurate simulator~\cite{gem} using L1I and L1D caches of 32KB each and an L2 (LLC) cache of 2MB. We added a timestamp and a per-hardware-context \emph{s-bit} to each cache line, which are manipulated as described in Section~\ref{sec:design}.  
The process context for a request packet in the cache is determined by the CR3 register within the simulator. Changes in the CR3 register are used to trigger the timestamp comparisons and the \emph{s-bit} saves and restores.  

%If an access results in a cache hit, the sbit is further checked to allow it to go through as a hit or a miss. If the sbit is reset, the request is sent down the memory hierarchy. The request is serviced with the existing cached data once the response is received without populating the cache, if the existing data is still valid.
%SD: seems redundant

Table~\ref{table:spectable} specifies the real and simulation system parameters used for the evaluation.
\begin{scriptsize}
\begin{table}[h!]
  \centering
\caption{Evaluation Setup}
  \label{table:spectable}
\begin{tabular}{ |c|c| } 
 \hline
 \hline
 \multicolumn{2}{| c |}{Real Processor}\\
 \hline
 Core & i7-7700, 3304.125 \\
 L1D, L1I, L2, LLC cache & 32K, 32K, 256K, 8192K\\
 \hline
  \multicolumn{2}{| c |}{gem5 Simulator}\\
 \hline
 Core & TimingSimpleCPU, 2GHz \\
 L1D, L1I, LLC cache & 32K, 32K, 2048K\\
 \hline
\end{tabular}
\end{table}
\end{scriptsize}

The following subsections present an analysis and evaluation of the security and the performance overheads of our timestamp-based defense on the gem5 simulator. 

\subsection{Security Analysis}

The attack depends on a fast reload due to another process. The attack can be broken if no process is allowed a cache hit due to another process. 
If the \emph{first access} by a process to an existing cache line is never a cache hit, the attacker remains oblivious of the data being cached beforehand and cannot learn if some shared data was accessed by another process. The second access is of no significance to the attacker. Allowing unaltered access beyond the \emph{first access} is sufficient to ensure security while not significantly compromising performance. A reuse attack in TimeCache can be prevented as follows:
\begin{enumerate}
    \item Attacker evicts a shared location
    \item Allows the victim process execution (caches shared data)
    \item Attacker accesses the shared location (does not experience cache hit due to victim)
\end{enumerate}
The attack does not go through. The additional information tracked for the defense includes timestamps and the \emph{s-bits}, are saved and restored by trusted software, and are protected from unprivileged access.

\subsubsection{Microbenchmark functionality evaluation}
In order to confirm the correct operation of the timestamp-based approach, we created a microbenchmark attack consisting of a pair of child and parent processes accessing a shared memory-mapped array of size equal to 256 cache lines. The parent process acts as the attacker, i.e., flushes the shared array and yields the processor. The victim's execution follows, where it writes a value repeatedly to the shared array. The parent process then wakes up and performs timed reads of the entire array. A hit is considered a successful attack. The attacker does not see any hit with our defense simulation enabled in gem5.
{\tt
\begin{lstlisting}
    if parent 
        flush shrd_mem;
        sleep;
        read shrd_mem; // cache hit 
    else
        read shrd_mem;
\end{lstlisting}
}

\subsubsection{Attacking RSA}

We use the {\tt flush+reload} technique to attack the GnuPG version of RSA, as described in the original paper~\cite{flushreload}. The attack was tested both on real hardware and the gem5 simulator, both running Linux. The attacker is an independent program, sharing the same machine and hence the caches.

On a real machine, we install a non-stripped GnuPG library and locate the offsets for the Square, Multiply, and Reduce functions. The shared library has the encryption algorithm for exponentiation, which performs a sequence of Square-Reduce-Multiply-Reduce for processing a key bit value 1 and a sequence of Square-Reduce when processing a clear bit. RSA encryption is an example where the control flow through the shared library is indexed using secret information, i.e., in this case, bit values from the secret key.

In the original attack, the attacker flushes the cache and then accesses the memory location for the Square, Multiply, and Reduce functions in a loop, using the time to  process a 1 or 0 bit coupled with whether or not accesses hit in the cache to extract information about the key being used. In our evaluation, we simplify the attack and assume a cache hit in the attacker process represents a successful attack.

We calculate the time required for a cached and uncached access on the experimental real machine and set that as the threshold for the cache hit. The attacker program is an independent program running a loop to flush and read memory. Reading the timestamps must be fenced/ordered with respect to the memory access being timed to avoid speculative loads. The attack goes through, i.e., the independent attacker program gets hits for its accesses as a simultaneously running victim process performs an encryption. We are able to launch the attack both on a real machine and in gem5 \emph{full-system} emulation mode.

Our defense in gem5 disallows any cache hit in the attacker process since the attacker's timed access is preceded by a flush. The defense allows a cache hit in a process only if it has suffered a cache miss for its \emph{first access}. Since the access after the flush to a cached data is the \emph{first access}, which is delayed, the attacker does not perceive a hit. This attack was the key demonstration for the {\tt flush+reload} attack and our defense successfully breaks the attack.

\subsubsection{S-bits Do Not Introduce Additional Side Channels}
The additional \emph{s-bits} do not introduce additional side channels for the following reasons:
\begin{itemize}
    \item they are not shared across processes either in hardware or software, are process specific and backed and restored with process context
    \item the \emph{s-bit} lookup does not depend on any secret, happens on every access, i.e. there is no information flow from process specific secret to the \emph{s-bits}
    \item \emph{s-bit} update happens at context switch and is a constant time operation
\end{itemize}

\subsection{Performance Evaluation}

\subsubsection{First Access Delay}
%\textcolor{pink}{We evaluate the performance overhead of our first-access delay mechanism by running SPECCPU2006 benchmarks in system-emulation mode. The benchmarks are run for 1 billion instructions.}
We evaluate the performance overhead of our first-access delay mechanism by running SPEC2006 and PARSEC in full system simulation of gem5.

\begin{figure*}[t!]
\caption{SPEC2006 performance overhead due to delayed first accesses; The average overhead is 1.5\% for two instance of same or different benchmarks on a single core.}
\label{specperf}
\centering
\includegraphics[width=0.85\textwidth]{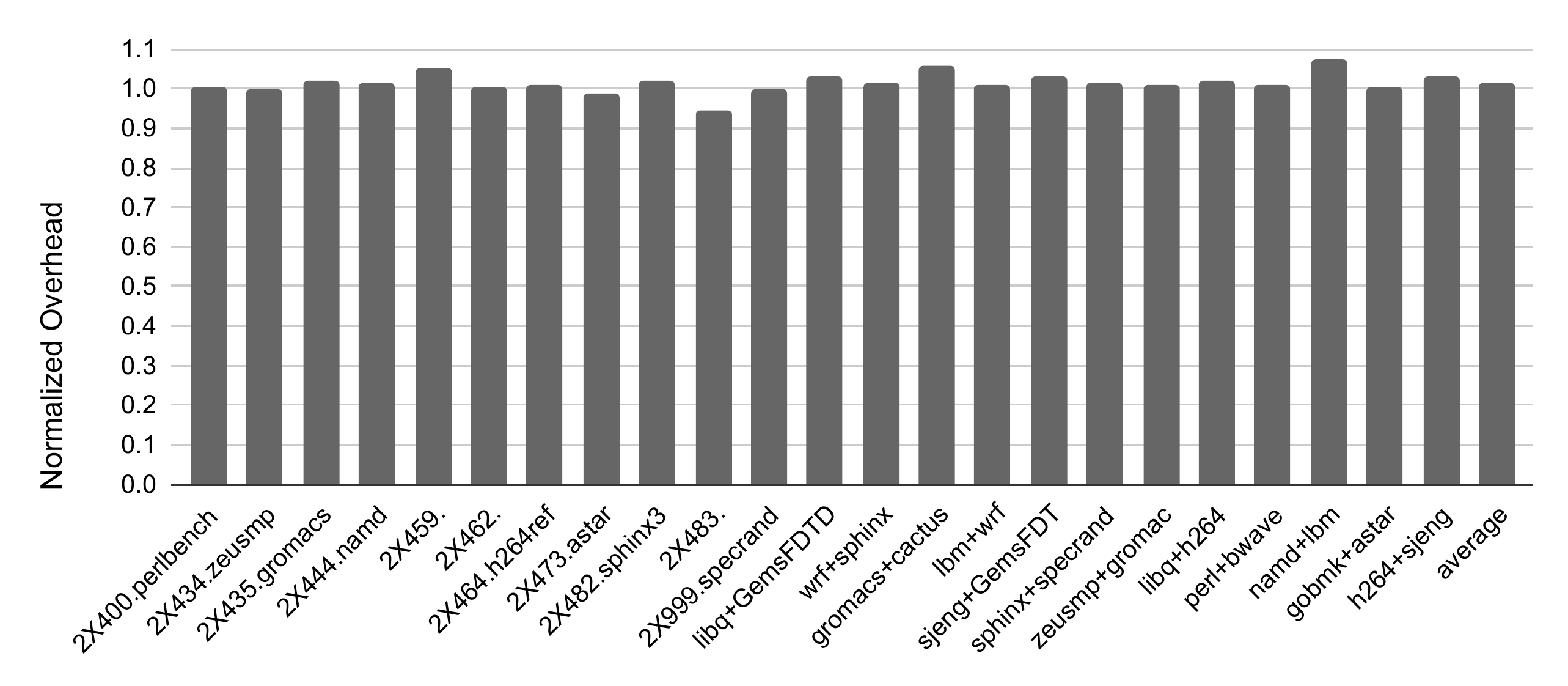}
\end{figure*}

\begin{figure}[tb]
\caption{PARSEC benchmark}
\label{parsec}
\centering
\includegraphics[width=0.45\textwidth]{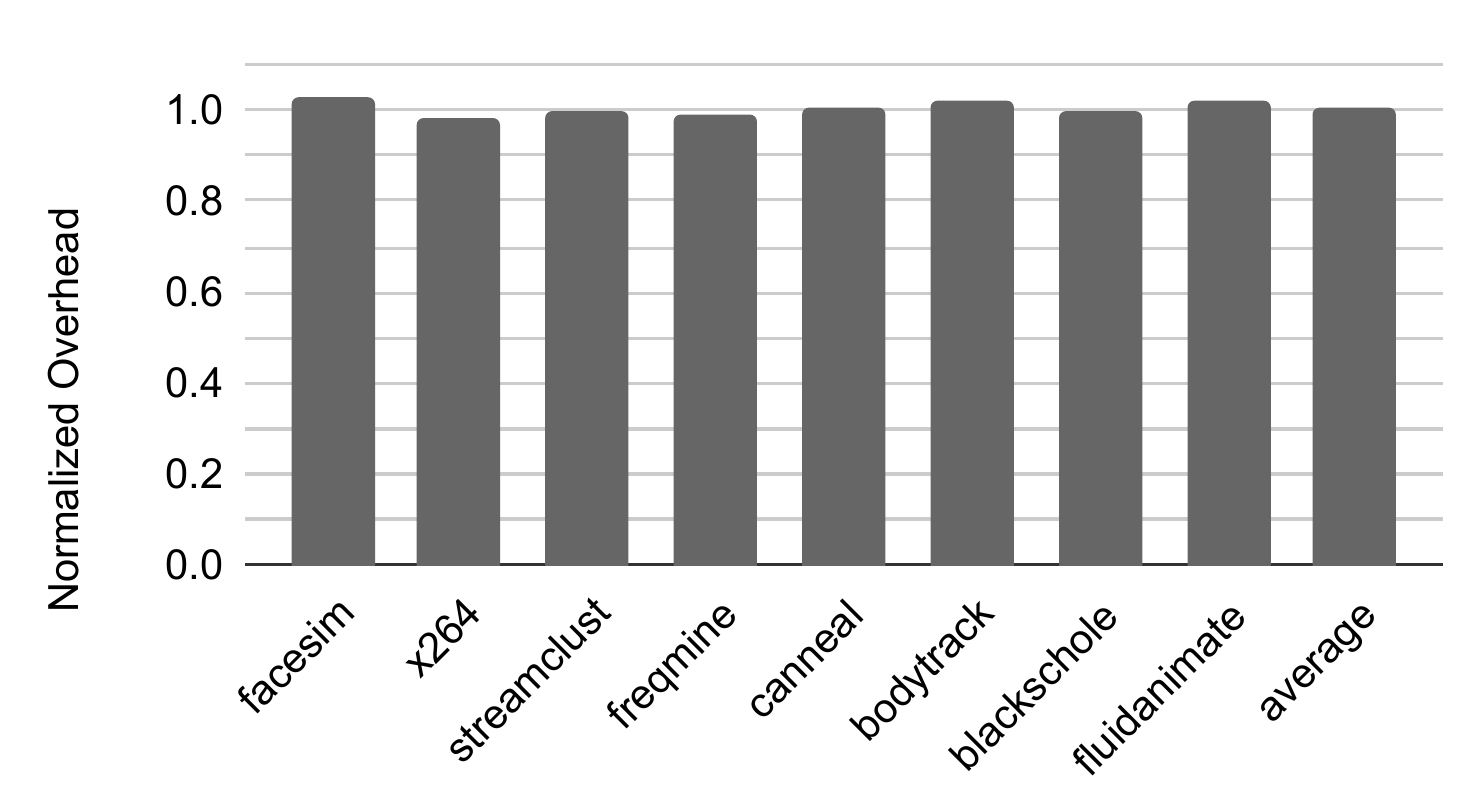}
\end{figure}

The performance overhead in Figure~\ref{specperf} is a graphical representation of the overheads due to misses on first accesses 
by running two instances of SPEC2006 benchmarks on a single shared core. When running 
two instances of a single benchmark, 
%\textcolor{pink}{Note that several of these benchmarks create multiple processes.} 
the number of first accesses is impacted by sharing both benchmark-specific code and shared libraries in the shared caches while context switching across these processes.  
We also  run a combination of different benchmarks on a single core, where the shared access is limited to shared libraries.  
%\textcolor{pink}{Figure~\ref{perf2} shows normalized execution time when running two different benchmarks simultaneously on two different cores}. 
The average overhead with two benchmark workloads is 
1.5\%. The overlap in common accesses and hence first accesses are limited to shared libraries with different workloads.
%\textcolor{pink}{ occupy the last-level cache.} 
The average impact on the execution time of PARSEC benchmark is 1.3\% as represented in Figure~\ref{parsec}.
The exact overheads and the change in the number of misses per thousand instructions (MPKI) for the last level of cache is presented in the Table~\ref{table:perftable}. The increase in MPKI is minimal, which explains the low overhead. 
%\textcolor{pink}{The lower miss rates with TimeCache are due to the increased number of total accesses to the LLC cache from the extra first access misses.}

\begin{figure}[tb]
\caption{Delayed Access MPKI at Each Cache Level}
\label{fraction}
\centering
\includegraphics[width=0.5\textwidth]{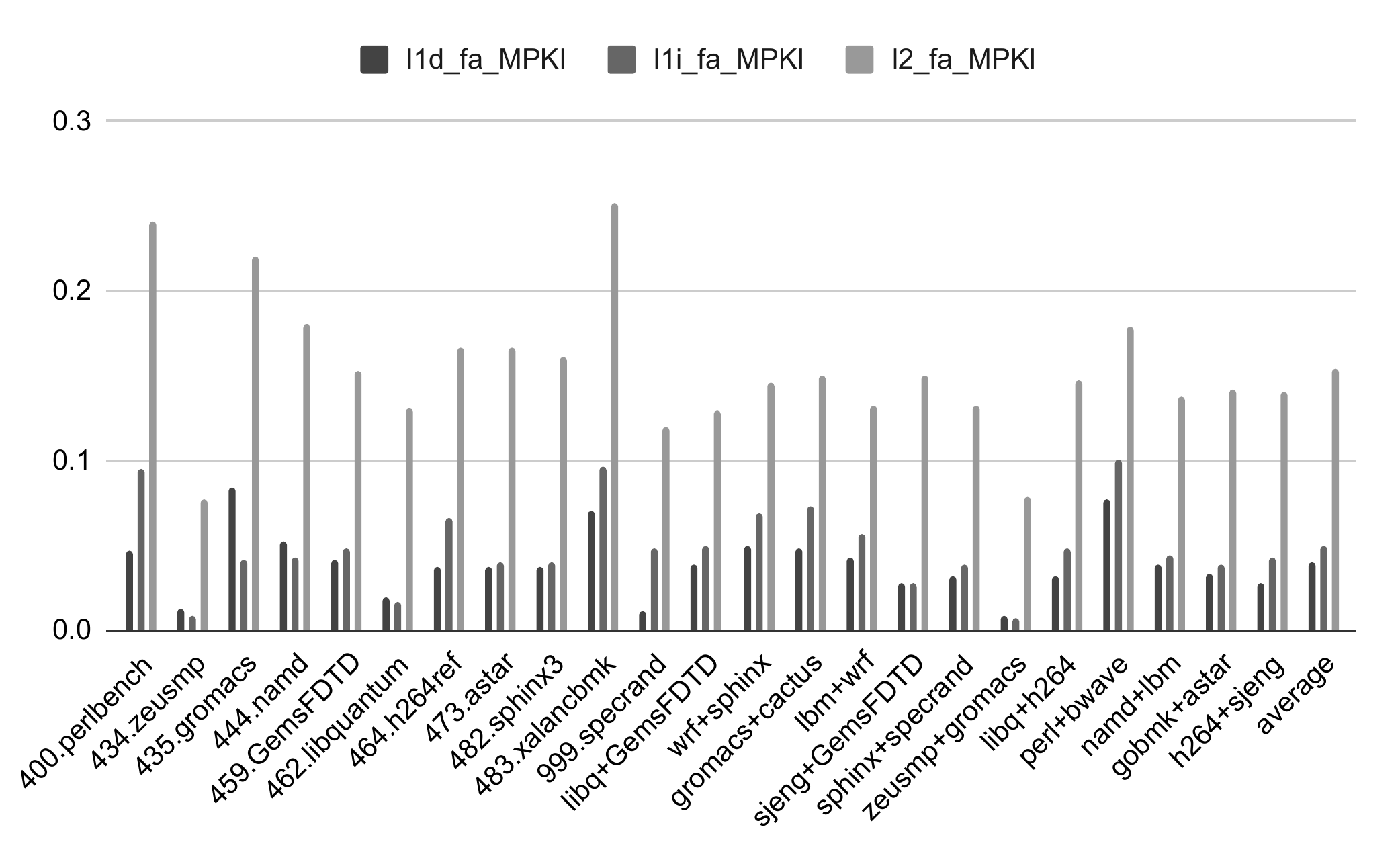}
\end{figure}

Figure~\ref{fraction} shows the ratio of \emph{first accesses} to the total number of accesses at each cache level for different benchmarks. 
%\textcolor{pink}{As can be seen, mcf, omnetpp, and perlbench have a higher fraction of first access misses, resulting in higher performance overhead as seen in Figure~\ref{perf1}. On the other hand, libquantum and wrf have a very low fraction of first access misses, so their performance is not impacted by the defense.} 
An interesting observation is that both 459.GemsFDTD and 462.libquantum have higher fractions of first access misses in the last-level cache when run individually.  
%\textcolor{pink}{and high overhead when run together}. 
However, when run together their effective first access misses are lower because of cache contention, resulting in a lower performance penalty. 
%\textcolor{pink}{Both astar\_perl and mcf\_gromac} 
Similarly 400.perlbench and 435.gromacs also have lower effective first access misses due to capacity evictions when sharing the cache with 410.bwaves and 434.zeusmp and hence lower overhead.

%\textcolor{pink}{The benchmarks that have several processes accessing shared libraries have a higher overhead due to delayed accesses since the caching benefits are not fully leveraged from one process to another. Thus, benchmarks like perlbench and omnetpp, which simulate network workloads by spawning several processes and using shared libraries among the processes have a slowdown larger than single process benchmarks like libquantum.}

\begin{scriptsize}
\begin{table}[h!]
  \centering
\caption{SPEC2006 and PARSEC Execution Time Overhead, 2MB LLC MPKI }
  \label{table:perftable}
\begin{tabular}{ |c|c|c|c| } 
 \hline
 Workload & Overhead & \thead{MPKI L2\\ Baseline}& \thead{MPKI L2\\TimeCache} \\
  \hline
 \hline
2X400.perlbench&1.003 &0.888&	1.131\\
2X434.zeusmp	& 1.000&	0.0323	&0.113\\
2X435.gromacs	& 1.022&	0.333	&0.553\\
2X444.namd	&1.016 &	0.180	&0.362\\
2X459.GemsFDTD &1.049 &		0.0766&	0.231\\
2X462.libquantum	&1.003 &	4.949	&5.082\\
2X464.h264ref	& 1.009&	2.687&	2.853\\
2X473.astar	&0.987 &	0.996&	1.161\\
2X482.sphinx3	& 1.021&	0.380&	0.540\\
2X483.xalancbmk &0.946  &		0.620&	0.870\\
2X999.specrand	& 1.001&	0.009&	0.130\\
libq+GemsFDTD	&1.029 &	2.431&	2.502\\
wrf+sphinx	& 1.015&	3.477	&3.571\\
gromacs+cactus &1.056 &		5.615	&6.434\\
lbm+wrf	& 1.008&	11.692&	11.638\\
sjeng+GemsFDTD	& 1.031&	6.171&	6.512\\
sphinx+specrand	&1.013 &	0.235&	0.371\\
zeusmp+gromacs	& 1.012&	0.423	&0.503\\
libq+h264	& 1.018&	3.869	&4.015\\
perl+bwave	&1.007 &	8.570&	8.427\\
namd+lbm	& 1.076&	6.413&	6.527\\
gobmk+astar	&1.006 &	6.287&	6.446\\
h264+sjeng	& 1.029&	7.974	&8.328\\
  \hline
average &1.015 &3.231	&3.40 \\
  \hline
facesim	&1.039&	0.484&	0.756\\
x264	&0.989&	1.670&	1.778\\
streamcluster&	1.011&	0.096&	0.10145\\
freqmine&0.994&	0.350&	0.423\\
canneal	&1.007&	6.575&	6.652\\
bodytrack&	1.031&	1.113&	1.409\\
blackscholes&	1.002&	3.363&	3.594\\
fluidanimate&	1.028&	1.691&	1.786\\
\hline
average	&1.012&	1.918&	2.062\\
  \hline
\end{tabular}
\end{table}
\end{scriptsize}

\subsubsection{LLC Size Sensitivity Analysis}
\begin{figure}[b]
\caption{Sensitivity Analysis} 
\label{sensitivity}
\centering
\includegraphics[width=0.35\textwidth]{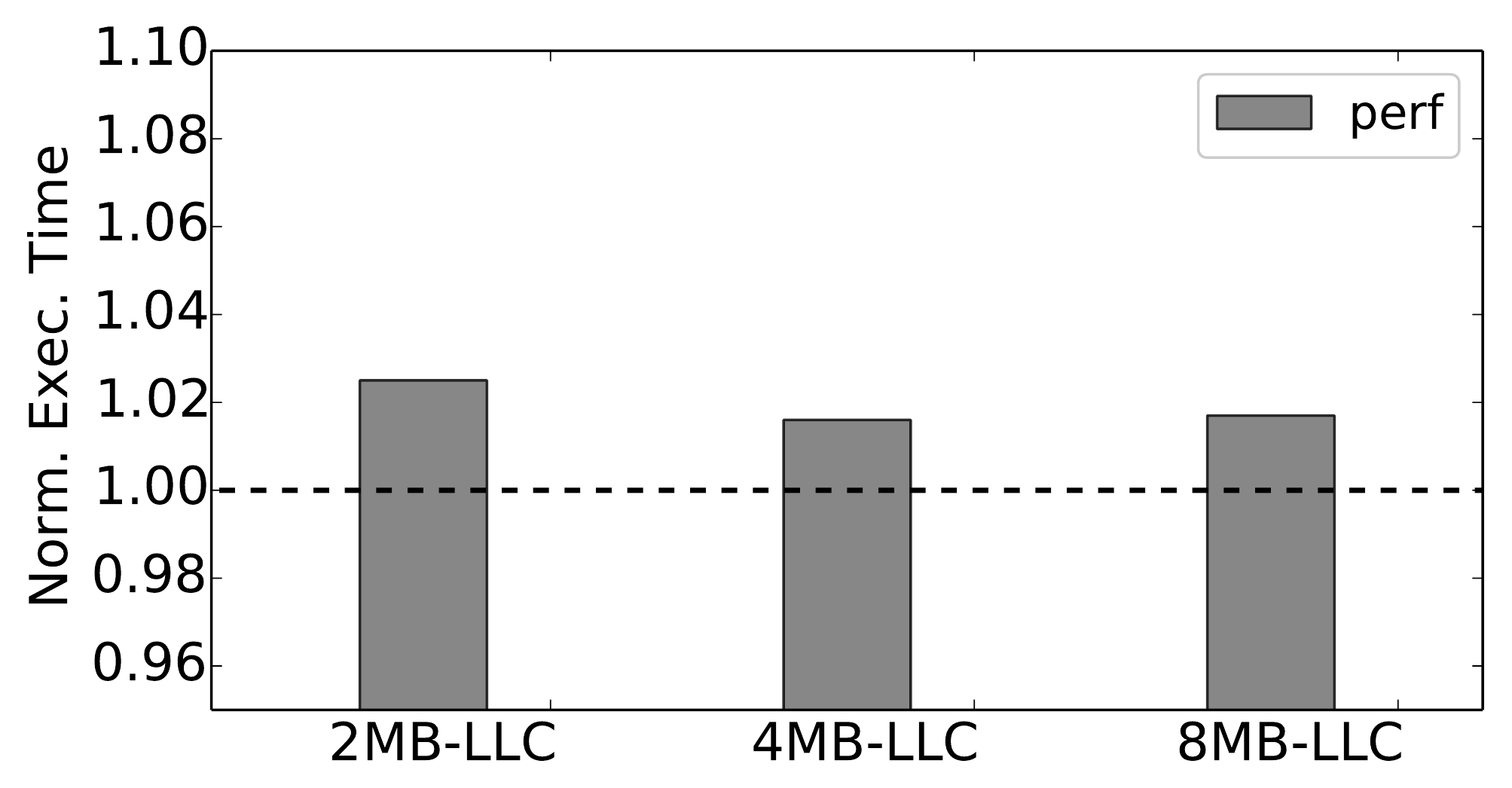}
\end{figure}

To analyze the sensitivity of our design to cache size, we evaluate the performance overhead with different LLC sizes for the single benchmark/single core tests (Figure~\ref{sensitivity}). Since the bigger caches are expected to have lower eviction rates for the same workload, there are effectively fewer first accesses, resulting in a smaller additional delay. Hence, we see the performance overhead in bigger caches to be smaller. Our analysis with 2MB, 4MB, and 8MB LLC sizes has an average 
performance overhead of 2.5\%, 1.6\%, and 1.7\% The ratio of first access miss to the overall miss also varies inversely with the cache size. The average first access miss to overall miss for the 2MB, 4MB, and 8MB cache sizes is 40\%, 10\% and 8\% respectively. These numbers indicate that the defense scales well with larger caches.

\subsection{Space Overhead, Timestamp Rollover, and Scaling}
\label{rollover}
The increase in area due to the additional hardware is primarily due to the separate SRAM array of timestamps and \emph{s-bits}, and the comparison logic.
This separate SRAM array uses 8-T rather then 6-T cells and also includes an additional set of sense-amps and bit-line drivers.
The other components required are the timestamp comparison logic at each bit-line peripheral, consisting of 2 latches and 2 3-input {\tt and} gates and a shift register to hold Ts.

In our evaluation, we use 32-bit timestamps to keep the area overhead low. The number of bits used for the timestamp counter has an impact on the frequency of timestamp rollover and is a parameter that can be tuned by the chip maker.  A timestamp rollover can result in an additional miss after $2^{32}$ cycles depending on the \emph{Ts} of the process. We illustrate the correctness of operation  using 2 decimal digits of precision \emph{Tc}, resulting in a rollover every 100 cycles for the purpose of illustration.

\begin{enumerate}
    \item Processes that preempt before and resume after rollover
    \begin{itemize}
        \item The rollover is detected by comparing \emph{Ts} and current time (e.g., 98 and 105).
        \item Since there can be newer unaccessed cache lines with rolled over (smaller) \emph{Tc} (e.g., 1(03))
        \item We reset all \emph{s-bits} when rollover is detected after a process resumes 
    \end{itemize}
    \item Processes that are running during a rollover
    \begin{itemize}
        \item No action is need while the process is running as the s-bits are up-to-date
        \item Assuming no rollover from \emph{Ts} and current time (e.g., 102 and 105)
        \item when the process resumes, since there can be older cache lines with bigger \emph{Tc} (78), unnecessary resets will occur, but correctness of operation is maintained.
    \end{itemize}
\end{enumerate}

Thus, cache line timestamp rollover can result in additional misses but retains correctness of the defense.
% These are the 2 cases to consider for a rollover (consider a 2 digit decimal counter). 
%
%1. processes that got pre-empted before rollover but resume after rollover (Ts 98).
%
%    When this process resumes, it can see cachelines with Tc 78 and Tc 102
%    Comparison would treat both 78 and 1(02) as smaller and will not reset the sbit, whereas the sbit should be reset for 1(02).
%    For correctness, rollover can be detected by the larger Ts and sbits reset on rollover.
%
%2. processes that were running during rollover, pre-empted and resumed later (Ts 107).
%
%    When this process resumes, cache lines may have timestamps Tc 78, Tc 1(08).
%    Comparison would treat 1(08) and 78 as greater than, where 78 would result in an unnecessary reset
%    For performance, we can either periodically flush or assume Tc will not remain 78 for long due to cache misses. 

An \emph{s-bit} is required per hardware context that shares the cache for each cache line.
% The sbits add to the area overhead by 0.005\% ((2/512)*(8/6)) for a single core and 0.02\% ((8/512)*(8/6)) for a 4-core chip, assuming 2 hyper-threads per core and 8T vs. 6T cells.
%The number of \emph{sbits} is therefore proportional to the number of hardware contexts sharing the cache. 
The total number of \emph{s-bits} can be significant for the LLC in server-class processors.  In order to keep the number of \emph{s-bits} low, design principles used for coherence directories could be applied, for example, limited pointers~\cite{agarwalsimoni88}.
The limited pointer~\cite{agarwalsimoni88} directory design work demonstrated empirically that applications typically share data across a few processors. Since pointers require log($n$) bits (for $n$ hardware contexts), keeping track of a limited number of sharers would reduce area overhead to O(log($n$)) as opposed to $n$ bits per cache line.
%This optimization can be useful when the cost of storing 1 bit per core exceeds the space r which is the case for shared LLC between more than 25 cores. Thus a processor with 256 cores would require 40 bits instead of 256. 

%~\cite{culler1999parallel}. 
%Due to the characteristics of t workload, there can be 5 to 6 sharers of a shared memory~\cite{culler1999parallel}. This suggests we could keep sbits for up to 5 sharers and still not hurt the performance, have reduced space requirement. This optimisation can be useful when the number of cores is more than 25, as the number of bits needed to represent a sharer is log(n).

\subsection{S-bits Save and Restore Overhead}
\label{sec:sbitoverhead}
When a process is resumed, the \emph{s-bits} and the \emph{Ts} that were saved for the process at the time of preemption must be restored. The overhead due to copying the \emph{s-bits} is low for small cache sizes. The entire \emph{s-bit} array for an L1 cache of size 64KB can be copied in 2 64-byte cache-line-size memory accesses. The overhead scales with the size of the cache. The copy can take 256 memory accesses for a last-level cache of size 8MB. The \emph{s-bits} can be read and written in parallel via the `regular' bit-line interface when a save or restore is required at a context switch. The save and restore is done to and from a kernel memory region reserved for the \emph{s-bits}, to which the process context points. 

\begin{scriptsize}
\begin{table}[h!]
  \centering
\caption{\emph{s-bit} size relative to the cache size}
\begin{tabular}{ |c|c|c| } 
 \hline
 cache size (B) & s-bits & accesses \\
 \hline
 64 & 1bit & 1 \\ 
 64K & 128B & 2\\ 
 256K & 512B & 8\\
 8M & 16KB & 256\\
 \hline
\end{tabular}
\end{table}
\end{scriptsize}

On an Intel i7-7700 processor operating at 3.6Ghz, the time to copy \emph{s-bits} for an 8MB size cache \emph{without} caching is 2.4 $\mu$ s. This is of comparable magnitude to a null context switch or system call.  A typical process time slice varies from 1 ms to several ms, so the 2.4 $\mu$s overhead is at most 0.24\% of the process run-time. An extra layer of buffering in hardware could allow the copy to be performed in parallel with the execution of the next process.

\section{Other Attacks On Shared Software}

\subsection{LRU Attack}
LRU attacks exploit the cache line replacement policy and depend on eviction set creation~\cite{lru}. They do not depend on shared software but can also be launched using shared software. The attack proceeds by creating an eviction set \emph{w}, and accessing a shared cache line \emph{l} followed by (\emph{w-1}) cache lines of the eviction set. After a time-lapse to allow the victim's execution, the attacker accesses the last element in the eviction set. The access replaces the oldest line, which is 1 of the  (\emph{w-1}) lines accessed by the attacker if the victim accesses \emph{l}. This attack, like the other contention attacks that require an  eviction set, can be prevented using randomizing caches.

\subsection{Coherence Attack}
Two types of attacks on shared memory due to coherence protocol have been identified in the literature~\cite{crosscoreAttack ,coherenceAttack}. One variant is know as \emph{invalidate+transfer}, where the attacker flushes a cache line, resulting in the cache line being flushed from all the private caches, and experiences a remote cache access latency on a subsequent load if the victim running on another processor accessed the same shared memory~\cite{crosscoreAttack}. The second variant exploits the difference in the access response time of \emph{Exclusive} and \emph{Shared} cached lines~\cite{coherenceAttack}.

We can prevent these types of attacks using TimeCache by waiting for a DRAM response even when the data is available in some remote cache or Last level Cache when the current context \emph{s-bit} is not set.

\subsection{FLush + Flush}
This attack uses the fact that the execution time of the `clflush' instruction depends on whether the data is available in the cache. The `clflush' instruction aborts early and takes less time if the data is not cached~\cite{flushflush}. If shared memory is accessed by the victim, the data is cached and rather than a fast reuse, the attacker can infer this by a second `clflush' taking longer. This form of attack can be prevented by making `clflush' a constant time instruction. One of the ways to do so is performing a dummy write back when the data is not cached.

\subsection{Evict + Time}
Evict+Time~\cite{primeprobe} is another contention-based attack, like the LRU attack, and does not depend on shared software, but can be launched using shared software.  The attacker evicts cache sets and checks if the evictions slows down the victim. A finer grain version of Evict+Time is possible on shared memory where the attacker flushes (using `clflush') a shared cache line and times the victim's execution. This attack remains noisy and less practical unless the attacker communicates with the victim to trigger and time a specific access.

\section{Related Work}
Existing solutions for protecting against cache side channel attacks that exploit shared code and data either resort to cache partitioning or remove timing information from the accesses. Both approaches incur significant overhead. 
\subsection{Cache Partitioning}~~
\label{subsec:secdcp}
While partitioning the cache can prevent multiple types of attacks, it comes at a performance cost due to reduced cache availability.
Statically partitioning caches causes significant performance deterioration as some parts of the cache become unavailable to other processes~\cite{static}, whereas dynamic cache partitioning can achieve lower overheads by reallocating space as needed. SecDCP~\cite{secdcp} is one such dynamic partitioning technique, which broadly categorizes applications as either `confidential' or `public' and prevents any information leakage from confidential applications to public applications, but allows information flow in the other direction. 
%SD: How?
%This mechanism is shown to have 12.5\% performance improvement over static cache partitioning~\cite{secdcp}. 
Although this dynamic cache partitioning technique performs better than its static counterpart, it provides a very coarse-grained security classification~\cite{secdcp}. Another dynamic cache partitioning technique utilizes page coloring to allocate pages to a secure domain~\cite{color,scolor}, but may incur significant copy costs for recoloring. DAWG partitions cache ways, supporting a maximum of 16 security domains at a time, and has an  associated performance overhead due to reduced cache availability of 4-12\%. PLCache defends against reuse attack by locking cache lines with process IDs to prevent their eviction by other processes~\cite{newcache}. It can be seen as a line-wise cache partitioning and since the locked lines are not available for eviction, there is a performance degradation of about 12\%.

%SD: why focus on the dynamic partitioning approach? What's important about it?

\subsection{Using Intel CAT-based Partitioning}~~
Both Catalyst~\cite{catalyst} and Apparition~\cite{dong2018shielding} have demonstrated the use of Intel's cache allocation technology (CAT) to achieve cache partitioning for mitigating cache side channels. 
Both Apparition and Catalyst disallow or do not protect against attacks using shared software. 
%SD: see above addition.
The performance of the systems depend on their ability to reassign caches to different applications and keep cache flushes to a minimum. Apparition~\cite{dong2018shielding} uses one Class of Service (CLOS) per application and flushes it across context switches.  
%The overhead can vary from being negligible to about 81\% or more~\cite{dong2018shielding}. 
Catalyst uses pinned pages to provide a solution suited to cloud service providers. 
%The performance overhead is about 6\%. 
This defense mechanism is suited to preventing cross-VM attacks and attacks targeted at the LLC, and is not suited to higher-level caches. 
The design further requires manually tagging pages that should be pinned or need to remain secure. 
%SD: how relevant are the performance numbers above?

\subsection{Removing Time \& Constant Time Algorithm}
The ability to time data accesses precisely can also be seen as the cause for side channel exploits. Taking this ability away from untrusted applications is not sufficient to prevent the attacks. There are several new techniques to obtain timestamps in up to microsecond granularity. These methods provide alternate timing primitives or recovery of clock resolution~\cite{fantastic} on systems that obfuscate time by reducing the clock resolution.

Other approaches to mitigating side channels in shared memory suggest program transformation for constant time implementation. These program transformations have impractical overhead due to making each critical access {\em O(n)}~\cite{rane15raccoon} and are not useful for large shared libraries.

%\subsection{Inferences}
%We have the following deductions from the survey of the past work :
\begin{comment}
We can draw the following inferences from the survey of the past work for the defense:
\begin{itemize}
    \item Static cache partitioning does not make efficient use of the entire cache, as a solution to which dynamic cache partitioning was introduced. Dynamic partitioning performs better than static partitioning~\cite{secdcp} but still has significant performance degradation.
    \item  Cache partitioning using Intel CAT provides a coarse partition granularity and has a restriction on maximum number of supported secure domains, defends only against attacks in LLC.
    \item Removing the capability to time transactions is not sufficient as there are several ways of deriving timing information~\cite{fantastic}.
    \item Constant time implementation remains a somewhat theoretical solution for most application because the overhead it adds can in effect render caches useless.
\end{itemize}
\end{comment}

\section{Discussion}

Sharing software is an important component of computing systems for efficiency and consistency. This work eliminates a channel for the leak of secret data via monitoring a victim's access to shared content using shared caches. In the absence of shared content, shared caches still allow a victim's access behavior to be monitored, but the information channel is far less accurate. In particular, a ``Prime+Probe'' attack fills (primes) an entire cache set, and infers the cache set accessed by the victim, based on whether the attacker's probe hits or misses. Proposed defenses for a ``Prime+Probe'' attack include a randomizing cache~\cite{ceasers}~\cite{newcache-micro-2016}. These defenses do not work for attacks against shared content, which provides a more accurate/less noisy channel of information.
TimeCache in conjunction with these defenses can provide a more complete defense. 
 
Other approaches to defending against more recent attacks like Spectre either stall execution, or make speculative instructions invisible to succeeding load requests~\cite{invisispec}~\cite{safespec}. They do not prevent non-speculative cache side channels. Speculative side channel attacks rely on conventional side channels for leaking speculatively loaded data to the attacker, i.e., the means of data leak is conventional side channels. Breaking conventional cache attacks, we also prevent speculative side channel leaks.

\section{Conclusion}
We have designed and evaluated a timestamp-based defense against timing side channel attacks that rely on reuse of shared software in caches to learn secret information. TimeCache works across context switches and prevents attacks from cross-core, same core, or SMT contexts, and at any level of cache, without the need for cache partitioning. 
To perform timestamp comparisons in parallel, we use an SRAM array that allows bit-serial, timestamp-parallel comparison with easy transposed access. 
We have evaluated the defense against microbenchmark attack programs and the classic {\tt flush+reload} attack using the gem5 simulator. On SPEC2006 and PARSEC, the performance overhead due to delaying the first accesses is 1.5\% and 1.2\% on average, and copying process-specific \emph{s-bits} adds at most 0.24\% even when there is a context switch every millisecond. Our defense against timing side channels through shared software retains the benefits of allowing processes to utilize the entire cache capacity of a shared cache and allows cache and memory pressure reduction through data deduplication and copy-on-write sharing.

\section{Acknowledgements}
This work was supported in part by National Science Foundation (NSF) Awards CNS-1618497 and CNS-1900803. We thank Sreepathi Pai for his feedback during early discussions of the ideas in this paper.

%%%%%%% -- PAPER CONTENT ENDS -- %%%%%%%%

%%%%%%%%% -- BIB STYLE AND FILE -- %%%%%%%%
\bibliographystyle{IEEEtranS}
\bibliography{cache}

% Generated by IEEEtranS.bst, version: 1.14 (2015/08/26)
\begin{thebibliography}{10}
\providecommand{\url}[1]{#1}
\csname url@samestyle\endcsname
\providecommand{\newblock}{\relax}
\providecommand{\bibinfo}[2]{#2}
\providecommand{\BIBentrySTDinterwordspacing}{\spaceskip=0pt\relax}
\providecommand{\BIBentryALTinterwordstretchfactor}{4}
\providecommand{\BIBentryALTinterwordspacing}{\spaceskip=\fontdimen2\font plus
\BIBentryALTinterwordstretchfactor\fontdimen3\font minus
  \fontdimen4\font\relax}
\providecommand{\BIBforeignlanguage}[2]{{%
\expandafter\ifx\csname l@#1\endcsname\relax
\typeout{** WARNING: IEEEtranS.bst: No hyphenation pattern has been}%
\typeout{** loaded for the language `#1'. Using the pattern for}%
\typeout{** the default language instead.}%
\else
\language=\csname l@#1\endcsname
\fi
#2}}
\providecommand{\BIBdecl}{\relax}
\BIBdecl

\bibitem{dedup2}
``Kernel samepage merging (memory deduplication),''
  \url{https://kernelnewbies.org/Linux_2_6_32#Kernel_Samepage_Merging_.28memory_deduplication.29},
  2017.

\bibitem{agarwalsimoni88}
A.~Agarwal, R.~Simoni, J.~Hennessy, and M.~Horowitz, ``An evaluation of
  directory schemes for cache coherence,'' in \emph{International Symposium on
  Computer Architecture (ISCA)}, Jun. 1988, pp. 280--289.

\bibitem{bitserial}
K.~E. Batcher, ``Bit-serial parallel processing systems,'' \emph{IEEE
  Transactions on Computers}, no.~5, pp. 377--384, 1982.

\bibitem{gem}
\BIBentryALTinterwordspacing
N.~Binkert, B.~Beckmann, G.~Black, S.~K. Reinhardt, A.~Saidi, A.~Basu,
  J.~Hestness, D.~R. Hower, T.~Krishna, S.~Sardashti, R.~Sen, K.~Sewell,
  M.~Shoaib, N.~Vaish, M.~D. Hill, and D.~A. Wood, ``The {GEM5} simulator,''
  \emph{SIGARCH Comput. Archit. News}, vol.~39, no.~2, pp. 1--7, Aug. 2011.
  [Online]. Available: \url{http://doi.acm.org/10.1145/2024716.2024718}
\BIBentrySTDinterwordspacing

\bibitem{fallout}
C.~Canella, D.~Genkin, L.~Giner, D.~Gruss, M.~Lipp, M.~Minkin, D.~Moghimi,
  F.~Piessens, M.~Schwarz, B.~Sunar \emph{et~al.}, ``Fallout: Leaking data on
  meltdown-resistant cpus,'' in \emph{Proceedings of the 2019 ACM SIGSAC
  Conference on Computer and Communications Security}, 2019, pp. 769--784.

\bibitem{sgxpectre}
G.~Chen, S.~Chen, Y.~Xiao, Y.~Zhang, Z.~Lin, and T.~H. Lai, ``{SGX}pectre
  attacks: Stealing {Intel} secrets from {SGX} enclaves via speculative
  execution,'' \emph{arXiv preprint arXiv:1802.09085}, 2018.

\bibitem{dong2018shielding}
X.~Dong, Z.~Shen, J.~Criswell, A.~L. Cox, and S.~Dwarkadas, ``Shielding
  software from privileged side-channel attacks,'' in \emph{27th $\{$USENIX$\}$
  Security Symposium ($\{$USENIX$\}$ Security 18)}, 2018, pp. 1441--1458.

\bibitem{neural}
C.~Eckert, X.~Wang, J.~Wang, A.~Subramaniyan, R.~Iyer, D.~Sylvester, D.~Blaauw,
  and R.~Das, ``Neural cache: Bit-serial in-cache acceleration of deep neural
  networks,'' in \emph{Proceedings of the 45th Annual International Symposium
  on Computer Architecture}.\hskip 1em plus 0.5em minus 0.4em\relax IEEE Press,
  2018, pp. 383--396.

\bibitem{flushflush}
D.~Gruss, C.~Maurice, K.~Wagner, and S.~Mangard, ``Flush+ flush: a fast and
  stealthy cache attack,'' in \emph{International Conference on Detection of
  Intrusions and Malware, and Vulnerability Assessment}.\hskip 1em plus 0.5em
  minus 0.4em\relax Springer, 2016, pp. 279--299.

\bibitem{templateattack}
D.~Gruss, R.~Spreitzer, and S.~Mangard, ``Cache template attacks: Automating
  attacks on inclusive last-level caches,'' in \emph{24th $\{$USENIX$\}$
  Security Symposium ($\{$USENIX$\}$ Security 15)}, 2015, pp. 897--912.

\bibitem{games}
D.~Gullasch, E.~Bangerter, and S.~Krenn, ``Cache games--bringing access-based
  cache attacks on aes to practice,'' in \emph{2011 IEEE Symposium on Security
  and Privacy}.\hskip 1em plus 0.5em minus 0.4em\relax IEEE, 2011, pp.
  490--505.

\bibitem{wei-ming92}
W.-M. Hu, ``Lattice scheduling and covert channels,'' in \emph{Proceedings 1992
  IEEE Computer Society Symposium on Research in Security and Privacy}.\hskip
  1em plus 0.5em minus 0.4em\relax IEEE, 1992, p.~52.

\bibitem{crosscoreAttack}
G.~Irazoqui, T.~Eisenbarth, and B.~Sunar, ``Cross processor cache attacks,'' in
  \emph{Proceedings of the 11th ACM on Asia conference on computer and
  communications security}, 2016, pp. 353--364.

\bibitem{dedup1}
\BIBentryALTinterwordspacing
K.~Jin and E.~L. Miller, ``The effectiveness of deduplication on virtual
  machine disk images,'' in \emph{Proceedings of SYSTOR 2009: The Israeli
  Experimental Systems Conference}, ser. SYSTOR ’09.\hskip 1em plus 0.5em
  minus 0.4em\relax New York, NY, USA: Association for Computing Machinery,
  2009. [Online]. Available: \url{https://doi.org/10.1145/1534530.1534540}
\BIBentrySTDinterwordspacing

\bibitem{safespec}
K.~N. Khasawneh, E.~M. Koruyeh, C.~Song, D.~Evtyushkin, D.~Ponomarev, and
  N.~Abu-Ghazaleh, ``Safespec: Banishing the spectre of a meltdown with
  leakage-free speculation,'' \emph{arXiv preprint arXiv:1806.05179}, 2018.

\bibitem{dawg}
V.~Kiriansky, I.~Lebedev, S.~Amarasinghe, S.~Devadas, and J.~Emer, ``Dawg: A
  defense against cache timing attacks in speculative execution processors,''
  in \emph{2018 51st Annual IEEE/ACM International Symposium on
  Microarchitecture (MICRO)}.\hskip 1em plus 0.5em minus 0.4em\relax IEEE,
  2018, pp. 974--987.

\bibitem{spectre}
P.~Kocher, D.~Genkin, D.~Gruss, W.~Haas, M.~Hamburg, M.~Lipp, S.~Mangard,
  T.~Prescher, M.~Schwarz, and Y.~Yarom, ``Spectre attacks: Exploiting
  speculative execution,'' \emph{arXiv preprint arXiv:1801.01203}, 2018.

\bibitem{spectreRSB}
\BIBentryALTinterwordspacing
E.~M. Koruyeh, K.~N. Khasawneh, C.~Song, and N.~Abu-Ghazaleh, ``Spectre
  returns! speculation attacks using the return stack buffer,'' in \emph{12th
  {USENIX} Workshop on Offensive Technologies ({WOOT} 18)}.\hskip 1em plus
  0.5em minus 0.4em\relax Baltimore, MD: {USENIX} Association, Aug. 2018.
  [Online]. Available:
  \url{https://www.usenix.org/conference/woot18/presentation/koruyeh}
\BIBentrySTDinterwordspacing

\bibitem{ghostrider}
C.~Liu, A.~Harris, M.~Maas, M.~Hicks, M.~Tiwari, and E.~Shi, ``Ghostrider: A
  hardware-software system for memory trace oblivious computation,'' \emph{ACM
  SIGPLAN Notices}, vol.~50, no.~4, pp. 87--101, 2015.

\bibitem{newcache-micro-2016}
F.~{Liu}, H.~{Wu}, K.~{Mai}, and R.~B. {Lee}, ``Newcache: Secure cache
  architecture thwarting cache side-channel attacks,'' \emph{IEEE Micro},
  vol.~36, no.~5, pp. 8--16, Sep. 2016.

\bibitem{catalyst}
F.~Liu, Q.~Ge, Y.~Yarom, F.~Mckeen, C.~Rozas, G.~Heiser, and R.~B. Lee,
  ``Catalyst: Defeating last-level cache side channel attacks in cloud
  computing,'' in \emph{2016 IEEE international symposium on high performance
  computer architecture (HPCA)}.\hskip 1em plus 0.5em minus 0.4em\relax IEEE,
  2016, pp. 406--418.

\bibitem{llcpractical}
F.~Liu, Y.~Yarom, Q.~Ge, G.~Heiser, and R.~B. Lee, ``Last-level cache
  side-channel attacks are practical,'' in \emph{2015 IEEE Symposium on
  Security and Privacy}.\hskip 1em plus 0.5em minus 0.4em\relax IEEE, 2015, pp.
  605--622.

\bibitem{winter}
M.~Mushtaq, M.~A. Mukhtar, V.~Lapotre, M.~K. Bhatti, and G.~Gogniat, ``Winter
  is here! a decade of cache-based side-channel attacks, detection \&
  mitigation for rsa,'' \emph{Information Systems}, p. 101524, 2020.

\bibitem{primeprobe}
D.~A. Osvik, A.~Shamir, and E.~Tromer, ``Cache attacks and countermeasures: the
  case of aes,'' in \emph{Cryptographers’ track at the RSA conference}.\hskip
  1em plus 0.5em minus 0.4em\relax Springer, 2006, pp. 1--20.

\bibitem{static}
\BIBentryALTinterwordspacing
D.~Page, ``Partitioned cache architecture as a {\.e}ide-channel defence
  mechanism,'' 2005. [Online]. Available:
  \url{http://citeseerx.ist.psu.edu/viewdoc/download?doi=10.1.1.460.9926&rep=rep1&type=pdf}
\BIBentrySTDinterwordspacing

\bibitem{ceaser}
M.~K. Qureshi, ``Ceaser: Mitigating conflict-based cache attacks via
  encrypted-address and remapping,'' in \emph{2018 51st Annual IEEE/ACM
  International Symposium on Microarchitecture (MICRO)}.\hskip 1em plus 0.5em
  minus 0.4em\relax IEEE, 2018, pp. 775--787.

\bibitem{ceasers}
------, ``New attacks and defense for encrypted-address cache,'' in
  \emph{Proceedings of the 46th International Symposium on Computer
  Architecture}.\hskip 1em plus 0.5em minus 0.4em\relax ACM, 2019, pp.
  360--371.

\bibitem{rane15raccoon}
A.~Rane, C.~Lin, and M.~Tiwari, ``Raccoon: Closing digital side-channels
  through obfuscated execution,'' in \emph{24th $\{$USENIX$\}$ Security
  Symposium ($\{$USENIX$\}$ Security 15)}, 2015, pp. 431--446.

\bibitem{rane16}
------, ``Secure, precise, and fast floating-point operations on x86
  processors,'' in \emph{25th $\{$USENIX$\}$ Security Symposium ($\{$USENIX$\}$
  Security 16)}, 2016, pp. 71--86.

\bibitem{fantastic}
M.~Schwarz, C.~Maurice, D.~Gruss, and S.~Mangard, ``Fantastic timers and where
  to find them: high-resolution microarchitectural attacks in javascript,'' in
  \emph{International Conference on Financial Cryptography and Data
  Security}.\hskip 1em plus 0.5em minus 0.4em\relax Springer, 2017, pp.
  247--267.

\bibitem{netspectre}
M.~Schwarz, M.~Schwarzl, M.~Lipp, J.~Masters, and D.~Gruss, ``Netspectre: Read
  arbitrary memory over network,'' in \emph{European Symposium on Research in
  Computer Security}.\hskip 1em plus 0.5em minus 0.4em\relax Springer, 2019,
  pp. 279--299.

\bibitem{singleton}
P.~Sharma and P.~Kulkarni, ``Singleton: system-wide page deduplication in
  virtual environments,'' in \emph{Proceedings of the 21st international
  symposium on High-Performance Parallel and Distributed Computing}, 2012, pp.
  15--26.

\bibitem{color}
J.~Shi, X.~Song, H.~Chen, and B.~Zang, ``Limiting cache-based side-channel in
  multi-tenant cloud using dynamic page coloring,'' in \emph{2011 IEEE/IFIP
  41st International Conference on Dependable Systems and Networks Workshops
  (DSN-W)}.\hskip 1em plus 0.5em minus 0.4em\relax IEEE, 2011, pp. 194--199.

\bibitem{checkmate}
C.~Trippel, D.~Lustig, and M.~Martonosi, ``Checkmate: Automated synthesis of
  hardware exploits and security litmus tests,'' in \emph{2018 51st Annual
  IEEE/ACM International Symposium on Microarchitecture (MICRO)}.\hskip 1em
  plus 0.5em minus 0.4em\relax IEEE, 2018, pp. 947--960.

\bibitem{ridl}
S.~Van~Schaik, A.~Milburn, S.~{\"O}sterlund, P.~Frigo, G.~Maisuradze,
  K.~Razavi, H.~Bos, and C.~Giuffrida, ``Ridl: Rogue in-flight data load,'' in
  \emph{2019 IEEE Symposium on Security and Privacy (SP)}.\hskip 1em plus 0.5em
  minus 0.4em\relax IEEE, 2019, pp. 88--105.

\bibitem{unveiling}
D.~Wang, A.~Neupane, Z.~Qian, N.~B. Abu-Ghazaleh, S.~V. Krishnamurthy, E.~J.
  Colbert, and P.~Yu, ``Unveiling your keystrokes: A cache-based side-channel
  attack on graphics libraries.'' in \emph{NDSS}, 2019.

\bibitem{secdcp}
Y.~Wang, A.~Ferraiuolo, D.~Zhang, A.~C. Myers, and G.~E. Suh, ``Secdcp: secure
  dynamic cache partitioning for efficient timing channel protection,'' in
  \emph{Proceedings of the 53rd Annual Design Automation Conference}.\hskip 1em
  plus 0.5em minus 0.4em\relax ACM, 2016, p.~74.

\bibitem{newcache}
Z.~Wang and R.~B. Lee, ``New cache designs for thwarting software cache-based
  side channel attacks,'' \emph{ACM SIGARCH Computer Architecture News},
  vol.~35, no.~2, pp. 494--505, 2007.

\bibitem{scattercache}
M.~Werner, T.~Unterluggauer, L.~Giner, M.~Schwarz, D.~Gruss, and S.~Mangard,
  ``Scattercache: thwarting cache attacks via cache set randomization,'' in
  \emph{28th $\{$USENIX$\}$ Security Symposium ($\{$USENIX$\}$ Security 19)},
  2019, pp. 675--692.

\bibitem{lru}
W.~Xiong and J.~Szefer, ``Leaking information through cache lru states,'' in
  \emph{2020 IEEE International Symposium on High Performance Computer
  Architecture (HPCA)}.\hskip 1em plus 0.5em minus 0.4em\relax IEEE, 2020, pp.
  139--152.

\bibitem{invisispec}
M.~Yan, J.~Choi, D.~Skarlatos, A.~Morrison, C.~Fletcher, and J.~Torrellas,
  ``Invisispec: Making speculative execution invisible in the cache
  hierarchy,'' in \emph{2018 51st Annual IEEE/ACM International Symposium on
  Microarchitecture (MICRO)}.\hskip 1em plus 0.5em minus 0.4em\relax IEEE,
  2018, pp. 428--441.

\bibitem{sharp}
M.~Yan, B.~Gopireddy, T.~Shull, and J.~Torrellas, ``Secure hierarchy-aware
  cache replacement policy (sharp): Defending against cache-based side channel
  attacks,'' in \emph{2017 ACM/IEEE 44th Annual International Symposium on
  Computer Architecture (ISCA)}.\hskip 1em plus 0.5em minus 0.4em\relax IEEE,
  2017, pp. 347--360.

\bibitem{coherenceAttack}
F.~Yao, M.~Doroslovacki, and G.~Venkataramani, ``Are coherence protocol states
  vulnerable to information leakage?'' in \emph{2018 IEEE International
  Symposium on High Performance Computer Architecture (HPCA)}.\hskip 1em plus
  0.5em minus 0.4em\relax IEEE, 2018, pp. 168--179.

\bibitem{flushreload}
\BIBentryALTinterwordspacing
Y.~Yarom and K.~Falkner, ``Flush+reload: A high resolution, low noise, l3 cache
  side-channel attack,'' in \emph{23rd {USENIX} Security Symposium ({USENIX}
  Security 14)}.\hskip 1em plus 0.5em minus 0.4em\relax San Diego, CA: {USENIX}
  Association, Aug. 2014, pp. 719--732. [Online]. Available:
  \url{https://www.usenix.org/conference/usenixsecurity14/technical-sessions/presentation/yarom}
\BIBentrySTDinterwordspacing

\bibitem{hdl}
D.~Zhang, Y.~Wang, G.~E. Suh, and A.~C. Myers, ``A hardware design language for
  timing-sensitive information-flow security,'' \emph{Acm Sigplan Notices},
  vol.~50, no.~4, pp. 503--516, 2015.

\bibitem{scolor}
X.~Zhang, S.~Dwarkadas, and K.~Shen, ``Towards practical page coloring-based
  multicore cache management,'' in \emph{Proceedings of the 4th ACM European
  conference on Computer systems}.\hskip 1em plus 0.5em minus 0.4em\relax ACM,
  2009, pp. 89--102.

\bibitem{PaaS}
\BIBentryALTinterwordspacing
Y.~Zhang, A.~Juels, M.~K. Reiter, and T.~Ristenpart, ``Cross-tenant
  side-channel attacks in paas clouds,'' in \emph{Proceedings of the 2014 ACM
  SIGSAC Conference on Computer and Communications Security}, ser. CCS
  ’14.\hskip 1em plus 0.5em minus 0.4em\relax New York, NY, USA: Association
  for Computing Machinery, 2014, p. 990–1003. [Online]. Available:
  \url{https://doi.org/10.1145/2660267.2660356}
\BIBentrySTDinterwordspacing

\end{thebibliography}
%%%%%%%%%%%%%%%%%%%%%%%%%%%%%%%%%%%%

\end{document}